\documentclass[a4paper,11pt]{article}
\pdfoutput=1 % if your are submitting a pdflatex (i.e. if you have
\usepackage{jheppub} % for details on the use of the package, please
\usepackage[T1]{fontenc} % if needed
\usepackage[usenames,dvipsnames]{xcolor}
\usepackage{feynmp-auto}

\newcommand{\half}{\frac{1}{2}}
\newcommand{\la}[1] {\left\langle #1 \right\rvert}
\newcommand{\ls}[1] {\left\lbrack #1 \bf \right\rvert}
\newcommand{\ra}[1] {\left\lvert #1 \right\rangle}
\newcommand{\rs}[1] {\left\lvert #1 \bf \right\rbrack}
\newcommand{\da}[1] {\left\langle #1 \right\rangle}
\newcommand{\ds}[1] {\left\lbrack #1 \bf \right\rbrack}
\newcommand{\dc}[1] {\left\{ #1 \bf \right\}}

\newcommand{\tl}[1] {\tilde{ #1 }}
\usepackage{verbatim}

\usepackage{amstext} % for \text macro
\usepackage{array}   % for \newcolumntype macro
\newcolumntype{L}{>{$}l<{$}} 

\title{Massive On-Shell Supersymmetric Scattering Amplitudes}

\author[a,b]{Aidan Herderschee,}
\author[a]{Seth Koren,}
\author[a]{and Timothy Trott}
\affiliation[a]{Department of Physics, University of California, \\
Santa Barbara, CA 93106, U.S.A.}
\affiliation[b]{Leinweber Center for Theoretical Physics, \\
Randall Laboratory of Physics, Department of Physics, \\
University of Michigan, Ann Arbor, MI 48109, USA}

\emailAdd{aidanh@umich.edu}
\emailAdd{koren@physics.ucsb.edu}
\emailAdd{ttrott@physics.ucsb.edu}

\abstract{We introduce a manifestly little group covariant on-shell superspace for massive particles in four dimensions using the massive spinor helicity formalism. This enables us to construct massive on-shell superfields and fully utilize on-shell symmetry considerations to derive all possible $\mathcal{N}=1$ three-particle amplitudes for particles of spin as high as one, as well as some simple amplitudes for particles of any spin. Throughout, the conceptual and computational simplicity of this approach is exhibited.}

\begin{document} 
\maketitle
\flushbottom

\section{Introduction}
\label{sec:intro}

The spinor helicity formalism has been a key ingredient in developing a purely on-shell formulation of $S$-matrix computations in four dimensions. This is because helicity spinors may be used as a complete description of the data of external scattering states (that is, their momentum and spin polarisation) without recourse to the unnecessary non-linear gauge redundancy of polarisations used in the Feynman rules. This can be coupled with on-shell methods, such as recursion and generalised unitarity, to perturbatively build the internal $S$-matrix structure out of on-shell units that bypass the need for the fictitious degrees of freedom that frequently arise in standard field-theoretic methods. However, the dream of a fully on-shell formulation of particle physics is far from realised.

Helicity spinors have been adapted to describe the kinematics of massive particles previously in \cite{Cohen:2010mi,Conde:2016vxs,Schwinn:2006ca,Kleiss:1985yh,Dittmaier:1998nn}. This generally involved decomposing time-like momenta into two null vectors and then proceeding with massless helicity spinors to describe each of these in some way. However, the element of arbitrariness in this decomposition often convoluted the method, as sought-after patterns could easily be obscured by an inappropriate choice. Furthermore, this choice of direction often involved a spurious breaking of Lorentz invariance in the amplitudes. 

An advance on this formalism was made in \cite{Arkani-Hamed:2017jhn}, where the spinors of the null vectors were organised into representations of the little group for massive momenta, $SU(2)$. This symmetry represents the redundancy in the spinor description of momentum, analogous to the $U(1)$ redundancy in the massless case. However, it may also be utilised to describe the polarisations of the external massive states, or better, to directly use symmetries to build amplitudes that have the required transformation properties of the external states under their individual little group rotations. While there was never a gauge ambiguity in the polarisations of massive particles in the Feynman rules, non-existent time-like components still source tension in a symmetric treatment of a $4$-vector description of these fundamentally $3$-vector objects (recent use in effective field theories was given in \cite{Shadmi:2018xan}). Massive and massless particles may thus be treated on equal footing within the spinor helicity formalism. 

Supersymmetry (SUSY) offers an idealisation that, in theories of massless particles, has enabled the utility of on-shell methods to be drastically extended. It is thus natural to look to supersymmetry as a testing grounds for on-shell methods for massive particles. We therefore here amalgamate the little group-covariant helicity spinors for massive particles with the formulation of an on-shell superspace in which external scattering states are grouped into supermultiplets without reference to an external spin direction. This makes the relations between the amplitudes imposed by the supersymmetric Ward identities (SWIs) transparent while simultaneously preserving the polarisation structures. See \cite{elvang2015scattering} for a review of on-shell superspace for massless particles. An on-shell superspace for massive particles was first constructed in \cite{Boels:2011zz} and we will rediscover their results along the way, albeit re-expressed in the covariant formalism. This helps to organise the amplitudes into Lorentz-covariant terms that are simpler to interpret, identify and construct.

After laying the foundation by writing the superalgebra in a little group covariant form and constructing covariant supermultiplets, we turn to $\mathcal{N}=1$ theories to exhibit the usage and utility of this formalism. We construct from first principles all possible three particle amplitudes, the most primitive on-shell scattering data, that are consistent with these symmetries and involve particle spins no greater than one. We also make some comments on how SUSY generally constrains interactions with higher-spin states. The on-shell supersymmetry allows us to simply catalogue the most general possible interactions given only the spectrum of a theory. It is also easy to further specialize by incorporating additional on-shell data such as the presence of a parity symmetry relating some component amplitudes to each other, or the absence of self-interactions for a vector in an Abelian theory. By studying the high energy or massless limits of superamplitudes, we may obtain the necessary dependence of the couplings on the masses of the external legs if the states are to be identified as elementary superfields.

This paper is structured as follows. In Section 2 we present the little group covariant on-shell SUSY algebra. This allows us in Section 3 to construct massive on-shell supermultiplets as coherent states of the supercharges in any reference frame. In Section 4 we discuss general features of superamplitudes and strategies for their construction, including the implementation of parity symmetry. We exhibit all of this technology in Section 5 to construct elementary three particle amplitudes for flat space $\mathcal{N}=1$ theories. We then conclude. 

\section{Little Group Covariant Superalgebra for Massive Particles}\label{allmassive}

The general super-Poincar\'{e} algebra extends the Poincar\'{e} algebra to a graded Lie algebra in $4$ dimensions through the introduction of $\mathcal{N}$ fermionic generators $Q_{\alpha A}$, $Q^\dagger_{\dot{\beta} B}$, where $\alpha, \dot{\beta}$ are $SL(2,\mathbb{C})$ indices and $A,B = 1 \dots \mathcal{N}$ count the number of left-handed spinor supersymmetry generators. The Lie brackets of the generators of supersymmetry - or supercharges - with the generators of translations $(P_\mu)$ and rotations/boosts $(M_{\mu\nu})$ are (following \cite{Srednicki:2007qs})
\begin{equation}
\begin{split}
[Q_{\alpha A},P^{\mu}]&=0, \\ 
[Q_{\alpha A},M^{\mu\nu}]&=\frac{i}{4} \epsilon^{\dot{\alpha}\dot{\beta}}( \sigma^\mu_{\alpha \dot{\alpha}} \sigma^\nu_{\beta \dot{\beta}} - \sigma^\nu_{\alpha \dot{\alpha}} \sigma^\mu_{\beta \dot{\beta}}) Q^\beta_A, \\
\{ Q_{\alpha  A},Q^{\dagger B}_{\dot{\beta}}\}&=-2 \delta_{A}^{B}(\sigma^{\mu}_{\alpha\dot{\beta}})P_{\mu}, \\
\{ Q_{\alpha A},Q_{\beta B}\}&=Z_{AB} \epsilon_{\alpha\beta},
\\
\{ Q^{\dagger A}_{\dot{\alpha}},Q^{\dagger B}_{\dot{\beta}}\}&=-Z^{AB}\epsilon_{\dot{\alpha}\dot{\beta}}.\label{FullSUSYalgebra}
\end{split}
\end{equation}

\noindent The automorphism group of the supercharges preserving the anticommutation relations is the $R$-symmetry group and will be discussed further in what follows. The `central charge' $Z_{AB}=-Z_{BA}=-(Z^{AB})^*$ is allowed for $\mathcal{N} > 1$ and typically breaks the $R$-symmetry to a subgroup. 

We will be interested in the construction of superamplitudes, which package together scattering data for entire representations of the super-Poincar\'{e} algebra. Before discussing this, we will first rewrite the superalgebra using the massive spinor helicity language. This allows the spinor indices to be stripped out of the supercharges, leaving an elegant, frame-independent formulation of the algebra from which massive representations can be simply constructed. This provides an aesthetic improvement over previous treatments \cite{Wess:1992cp} in addition to setting up our discussion of superamplitudes. 

In Appendix \ref{Review}, we present a lightning review of massive spinor helicity in which we also develop our conventions and provide relevant and useful identities. The reader can find there further introduction to the subject and the elementary mechanics which will not be remarked upon in the main text.

The external particles in a scattering amplitude are acted upon by the super-Poincar\'{e} generators as separate tensor factors of the scattering state. Each symmetry generator may therefore be represented on a scattering amplitude as the sum of its action on each external scattered particle. This will allow us to study symmetry generators and transformations on each leg separately. We will use spinor helicity variables to represent these generators, because these encapsulate the on-shell kinematic data for each leg. For massive particles, this means exhibiting the $SU(2)$ little group symmetry by expressing the symmetry generators acting on each particle $i$ in an appropriately covariant fashion. By construction, the momentum eigenvalue of particle $i$ is $p_i^{\dot{\alpha} \beta} = p_i^\mu \sigma_\mu^{\dot{\alpha} \beta} = \ra{i^I}^{\dot{\alpha}} \ls{i_I}^\beta$. We can likewise define on-shell, little group covariant supersymmetry generators for each leg by projecting the supercharges onto the spinors of a given particle
\begin{equation}
q^I_{i,A} = \frac{-1}{\sqrt{2}m_i} \ds{i^I Q_{i,A}}, \qquad q^{\dagger A}_{i,I} = \frac{1}{\sqrt{2}m_i} \da{i_I Q^{\dagger A}_{i}},\label{supercharges}
\end{equation}

\noindent where the factor of mass $m_i$ of the particle makes them dimensionless. Note that we are defining these operators as being restricted to single-particle momentum eigenspaces. For convenience, the inverse relations are given by
\begin{align}
Q_{i, \alpha A} = -\sqrt{2}\rs{i_I}_\alpha q^I_{i,A}\qquad Q^{\dagger A}_{i,\dot{\beta}} = \sqrt{2}q^{\dagger A}_{i,I} \la{i^I}_{\dot{\beta}}.\label{superchargesInv}
\end{align}
The factor of $1/\sqrt{2}$ is just a normalisation convention and has been chosen here so that the little group covariant supercharges satisfy the anticommutation relations
\begin{equation}
\dc{q_{i,A}^{I},q^{\dagger J,B}_{i}}=-\epsilon^{IJ}\delta_{A}^{B}, \qquad \dc{q_{i,A}^{I}, q_{i,B}^{J}}=-\epsilon^{IJ}\frac{Z_{i,AB}}{2m_i}, \qquad \dc{q_{i}^{\dagger I,A}, q_{i}^{\dagger J,B} }=\epsilon^{IJ}\frac{Z_i^{AB}}{2m_i}.\label{SUSYalgebra}
\end{equation}
These hold only on a particular single-particle momentum eigenspace, the labeling of which we leave implicitly subsumed in the particle label $i$. Here $Z_i$ is the particle's central charge. Also of note is that, as a result of the way the massive spinors transform under conjugation, $(q_{I,A})^\dagger=-q^{\dagger I,A}$ and $(q^I_A)^\dagger=q^{\dagger A}_I$. As usual, the $SU(2)$ little group indices may be raised and lowered using the Levi-Civita symbol. When the external legs are massless, the supercharges (\ref{superchargesInv}) become
\begin{align}
Q_{i, \alpha A} = -\sqrt{2}\rs{i}_\alpha q_{i,A}\qquad Q^{\dagger A}_{i,\dot{\beta}} = -\sqrt{2}q^{\dagger A}_{i} \la{i}_{\dot{\beta}}.\label{superchargesInvMassless}
\end{align}
The little group covariant supercharges satisfy the algebra
\begin{equation}
\dc{q_{i,A},q^{\dagger B}_{i}}=\delta_{A}^{B}\label{masslessSUSYalgebra}
\end{equation}
and the other anticommutators are zero.

The stripping of the helicity spinor effectively exchanges manifest chirality for manifest spin polarisation (of which helicity is often a natural and useful example). For massless states, these are identical and each chiral spinor supercharge can only either raise or lower a state's helicity. However, for massive states, the supercharges in the form of chiral spinors will do a superposition of both, for the usual reason that chirality and helicity/polarisation are no longer identical. The little group here describes the freedom in choosing a spin direction as a state label, which determines how the chiral spinor supercharges are decomposed into supercharges characterised by polarisation.

In the simple case in which all legs carry a single, electric central charge, $Z_{i,AB}=Z_i\,\Omega_{AB}$, where $Z_i\in\mathbb{R}$ while $\Omega_{AB}=-\Omega_{BA}$ is a symplectic $2$-form: 
\begin{equation}\label{eqn:centralcharge}
\Omega_{AB}=\begin{bmatrix}
 0 & -I \\ 
 I & 0
\end{bmatrix},
\end{equation}
where $0$ is the $\frac{\mathcal{N}}{2}\times\frac{\mathcal{N}}{2}$ zero matrix, while $I$ is the identity of the same size. Specifically for $\mathcal{N}=2$, $\Omega_{AB}=\epsilon_{AB}$, $Z_i$ may be complex (corresponding to two central charges) and this central extension is general. The supercharge labels $A,B$ give a manifest representation of a symmetry group that acts on $q^I_A$ (and on $q^{\dagger IA}$ in the conjugate representation) while preserving the algebra (\ref{SUSYalgebra}). If $Z_{AB}=0$, this would be $SU(\mathcal{N})$ (or $U(\mathcal{N})$), while for the central charge considered above, this would be broken to $USp(\mathcal{N})$. The symplectic $2$-form $\Omega_{AB}$ may then be used to convert $USp(\mathcal{N})$ tensor representations (such as the supercharges) into conjugate representations (i.e. raise and lower the explicit $R$-indices) in the way that the Levi-Civita tensor does for $SU(2)$. 

For $|Z_i|<2m_i$, the relations (\ref{SUSYalgebra}) may be simplified. Unlike for massless particle representations, the generators $q_{I,A}$ and their conjugates $q^{\dagger I,A}$ may mix because their index heights may be changed by $\epsilon_{IJ}$ and $Z_{AB}$. This allows for a rotation into a basis that canonicalises the anticommutators. This basis is given by 
\begin{equation}
\bar{q}^I_{i,A}=\frac{1}{\sqrt{D}}\left(q^I_{i,A}+\left(\frac{2m_i}{|Z_i|}\right)^2\left(1-\sqrt{1-\left(\frac{|Z_i|}{2m_i}\right)^2}\right)\frac{Z_{i,AB}}{2m_i}q^{\dagger B,I}_{i}\right).\label{SuperChargeRedef}
\end{equation}
where $D=2\Big(\Big(\frac{2m_i}{|Z_i|}\Big)^2-1\Big)\Big(1-\sqrt{1-\Big(\frac{|Z_i|}{2m_i}\Big)^2}\Big)$. The $\bar{q}_{I,A}$ and their conjugates then satisfy the anticommutation relations without a central charge:
\begin{equation}
\dc{\bar{q}_{i,A}^{I},\bar{q}^{\dagger J,B}_{i}}=-\epsilon^{IJ}\delta_{A}^{B}, \qquad \dc{\bar{q}_{i,A}^{I}, \bar{q}_{i,B}^{J}}=0, \qquad \dc{\bar{q}_{i}^{\dagger I,A}, \bar{q}_{i}^{\dagger J,B} }=0.\label{SUSYalgebraDiag}
\end{equation}

\noindent In such cases, representations of the supersymmetry algebra may be constructed with a structure identical to that of the case with $Z_i=0$, although such multiplets still carry central charge (and this would still appear in relating $\bar{q}_{i,A}^{I}$ and $\bar{q}_{i}^{\dagger I,A}$ to $Q_{\alpha A}$ and $Q^\dagger_{\dot{\beta}B}$ for these states). Henceforth, this redefinition of the particles' supercharges will be implicit in subsequent discussions of SUSY representations with central charges satisfying $|Z_i|<2m_i$ and the bars on the diagonalised supercharges will be omitted.

The relations (\ref{SUSYalgebraDiag}) illustrate the symplectic $R$-symmetry of the massive representation. While (\ref{SUSYalgebra}) has a manifest $USp(\mathcal{N})$ $R$-symmetry, because the supercharges can mix with their conjugates while preserving the $SU(2)$ little group symmetry, the full $R$-symmetry group is actually determined by all of the automorphisms that preserve the anticommutation relations (\ref{SUSYalgebraDiag}). Grouping the supercharges into a $2\mathcal{N}$ length vector $\mathbf{q}_{i,a}^I=(\bar{q}_{i,A}^I,\bar{q}^{\dagger I,B}_{i})$, where $a=A$ for $a\leq \mathcal{N}$ and $a=B+\mathcal{N}$ for $a>\mathcal{N}$, (\ref{SUSYalgebraDiag}) may be combined into the relation
\begin{equation}
\dc{\mathbf{q}^I_{i,a},\mathbf{q}^J_{i,b}}=\epsilon^{IJ}\Omega_{ab}.
\end{equation}
Here $\Omega_{ab}$ is a $2\mathcal{N}\times 2\mathcal{N}$ symplectic $2$-form. Thus the anticommutator is effectively itself a symplectic $2$-form and the $R$-symmetry is enhanced to $USp(2\mathcal{N})$ \cite{Ferrara:1980ra}. However, it is often broken by interactions. The enlarged $R$-symmetry does not occur for massless representations of the SUSY algebra because the non-zero supercharges have definite opposite helicity and cannot mix. 

The case $|Z_i|=2m_i$ is the special BPS limit. This typically occurs for elementary particles which obtain mass through Higgsing of a vector multiplet \cite{Fayet:1978ig}. The redefinition of supercharges that give the canonical anticommutation relations described above fails in the BPS limit. This is because, for these representations, half of the number of supercharges are eliminated through the reality constraint
\begin{equation}
q_{i,IA}=\frac{-1}{2m_i}Z_{i,AB}q^{\dagger B}_{i,I}.
\label{bpslimit}
\end{equation}

\noindent The phase of $Z_i$ may be absorbed into a redefinition of the supercharge. Calling this time $\bar{q}_{i,IA}=q_{i,IA}e^{-i(\arg Z)/2}$, the BPS condition reduces to
\begin{equation}
\bar{q}^{IA}_{i}=-\bar{q}^{\dagger IA}_{i} \qquad \bar{q}_{i,IA}=-\bar{q}^\dagger_{i,IA}.\label{BPSLG}
\end{equation}
This condition again preserves the supersymmetry algebra. Clearly, BPS states are annihilated by the combination $\bar{q}^{IA}_{i}+\bar{q}^{\dagger IA}_{i}$. For the central charge considered above with $Z_{AB}\propto\Omega_{AB}$, the multiplet is $1/2$-BPS as it is annihilated by half of the supercharges. Configurations with multiple central charges are also possible in which some smaller fraction of supercharges annihilate the state.

The explicit $SU(\mathcal{N})$ symmetry of the SUSY algebra, which is broken to $USp(\mathcal{N})$ by the central charge of these massive single particle states, is therefore the massive $R$-symmetry group expected for a theory with half of the number of supersymmetries. A $1/2$-BPS state in $\mathcal{N}$-SUSY may be represented as a massive non-BPS state of $\mathcal{N}/2$-SUSY. For example, for the simplest spontaneous symmetry breaking pattern in $\mathcal{N}=4$ SYM, the massless $SU(4)$ $R$-symmetry is broken to $USp(4)$ when the central charge is generated. As the former is unbroken by dynamics and imposes stringent selection rules on scattering amplitudes at the origin of the moduli space, the latter should also be respected by the dynamics and organise the transition matrix structure away from the origin. See \cite{Herderschee:2019dmc} for further discussion. Further elaboration upon the representation of BPS states in scattering amplitudes has been recently made in \cite{Caron-Huot:2018ape}.

More generally, with more complicated configurations of active central charges than the simple case discussed above, for each $1/2$-BPS leg there is nevertheless an $SU(\mathcal{N})$ $R$-basis in which the central charge can be rotated into the form $Z_{i,AB}\propto\Omega_{AB}$. In such a basis, the representation of the leg's supercharges is just as described. However, as this basis is different for each leg, the linear combinations of supercharges that annihilate each state may differ by a $SU(\mathcal{N})$ rotation matrix, which must be accounted for when adding together the total supercharges. The $R$-symmetry group will also be broken further beyond $USp(\mathcal{N})$, but this will still be a symmetry restricted to the algebra of a single leg's supercharges. 

Finally, the BPS bound itself, $|Z_i|\leq 2m_i$, may be derived for these scattering states from the fact that the operator,
\begin{align}
\left(q_{i,IA}+\frac{1}{2m_i}Z_{i,AB}q_{i,I}^{\dagger B}\right)\left(q_{i,IA}+\frac{1}{2m_i}Z_{i,AC}q_{i,I}^{\dagger C}\right)^\dagger\label{BPSComb}
\end{align}
being a sum of squares, must have non-negative spectrum. Using the algebra (\ref{SUSYalgebra}), (\ref{BPSComb}) simplifies to $q_{i,IA}\left(q_{i,IA}\right)^\dagger\left(1-\frac{|Z_i|^2}{(2m_i)^2}\right)$. The BPS bound follows by simply requiring that this be non-negative.

\section{On-Shell Supermultiplets} \label{sec:multiplets}

We seek here to construct scattering amplitudes for supersymmetric theories, so need to understand the structure of supersymmetric scattering states. Scattering data is simplified considerably by the grouping of component states into coherent states of the supersymmetry algebra, known as `on-shell superfields'. For these scattering states we may describe their collective $S$-matrix entries using superamplitudes, which manifest both the supersymmetric Ward identities in a simple manner. 

For massless theories, on-shell superfields have been established as a convenient organisation of the representations \cite{Elvang:2011fx}. The on-shell superspace was first conceived of in \cite{Nair:1988bq} for $\mathcal{N}=4$ super Yang-Mills (SYM) and was employed later in \cite{Witten:2003nn} to formulate the supertwistor space representation of tree-level scattering amplitudes in these theories (it is worth noting that an off-shell superspace formulation of $\mathcal{N}=4$ SYM does not yet exist). In particular, for $\mathcal{N}=4$ SYM, it makes transparent the classification of the amplitudes into sectors of a fixed order of helicity violation, which close under both supersymmetry and $R$-symmetry \cite{ArkaniHamed:2008gz,Drummond:2008vq,Brandhuber:2008pf}. This enabled the formulation of the super-BCFW shift \cite{ArkaniHamed:2008gz,Brandhuber:2008pf} and the subsequent construction of all tree amplitudes and loop-level integrands in the limit of large gauge group dimension \cite{Drummond:2008cr,He:2010ju, ArkaniHamed:2010kv}, as well as the elucidation of the dual superconformal symmetry and dual twistor representations of amplitudes on complex projective space \cite{Drummond:2008vq}. See \cite{elvang2015scattering} for a review of these topics. Amplitudes in theories with fewer supersymmetries have also been formulated in an on-shell superspace in \cite{Elvang:2011fx}, where on-shell superfields for $\mathcal{N}<4$ massless theories were constructed. We refer the reader to these papers and the review for details of the construction of superamplitudes for massless theories, and now turn to the construction of massive supermultiplets. 

General on-shell superspaces for massive states have been previously developed in \cite{Boels:2011zz}. However, the manifestation of the massive little group for the external legs will allow us to improve upon the presentation of this exposition, as well as providing flexibility to choose a spinor basis best suited for the study of particular phenomena, such as high energy limits or complex momentum shifts. Much of the subsequent discussion here will parallel that of \cite{Boels:2011zz}, with the improved organisation offered by the little group. This has been utilised more recently for $\mathcal{N}=4$ super-Yang-Mills in \cite{Cachazo:2018} and will be elaborated upon in this context in \cite{Herderschee:2019dmc}.

From (\ref{SUSYalgebraDiag}), the massive supersymmetry algebra is that of $N$ fermionic oscillators, where $N=2\mathcal{N}$ if the representation is not BPS, but can be reduced by up to a factor of $1/2$ if shortened. Supermultiplets may be represented as coherent states which are eigenstates of $N$ `lowering operators'. To build these states we introduce Grassmann variables which transform in the little group of each particle $\eta^A_{i,I}$, as well as their conjugates $\eta^{\dagger I}_{i,A}$. The labels here match those of the supercharges $q^I_{i,A}$. In this section, we will restrict our attention to multiplets which are not BPS, in which case the oscillator index may be identified with a supercharge $R$-index (`$A$' in the symbols just introduced). Following the conventions of \cite{elvang2015scattering}, all particles will be represented as outgoing scattering states. We will reverse the heights of $R$-indices relative to this reference.

To ensure little group and $R$-covariance, either all of the $q_{i,IA}$ or all of the $q^{\dagger A}_{i,I}$ will be chosen as the lowering operators. These will have some Clifford vacuum states, $\la{\Omega}$ and $\la{\overline\Omega}$, which are annihilated by either set. Generally, any linear combination of $q_{i,IA}$ and $q^{\dagger A}_{i,I}$ for each such pair can be chosen as annihilation operators, the choice of which corresponds to the selection of a particular state in the multiplet as the Clifford vacuum in the superfield representation, but a choice that yields the most manifest symmetries is arguably desirable.

An entire supermultiplet may be encoded as a coherent state
\begin{equation}\label{coherent}
\la{\eta_i} = \la{\Omega} e^{q^{ I}_{i,A}\eta^A_{i,I} }
\end{equation} 
where $\eta^A_{i,I}$ are anticommuting Grassmann algebra generators. As is clear in this definition and will be made manifest below, the entire superfield transforms coherently under little group transformations with the same little group weight as the Clifford vacuum. The action of the supercharges on the states generalizes the action for massless particles described in \cite{elvang2015scattering}, where little group and $R$-indices of the supercharge must be tensored together and then decomposed. These are eigenstates of the annihilation operators, satisfying $\la{\eta_i}q^{\dagger A}_I= \la{\eta_i}(-\eta_{i,I}^A)$. The Grassmann Fourier transform may be used to define a basis of conjugate states. It is defined with its inverse respectively as:
\begin{equation}
\begin{split}
\la{\eta^{\dagger}} = \int \text{d}^{2\mathcal{N}}\eta \ e^{\eta^A_{I}\eta^{\dagger I}_A} \la{\eta}\qquad
\la{\eta} = \int \text{d}^{2\mathcal{N}}\eta^\dagger \  e^{\eta^{\dagger I}_A\eta_{I}^A} \la{\eta^{\dagger}}
\end{split}
\end{equation} 
The fact that both the two different $\eta$ and $\eta^\dagger$ representations for the same supermultiplet exist and are related by the Fourier transform will be useful in constraining the form of superamplitudes.

In the $\eta$ basis, the supercharges act as (assuming for simplicity the absence of central charges)
\begin{align}
\la{\eta_i}\da{\theta_A Q^{\dagger A}} = -\sqrt{2}\da{\theta_A i^I}\eta_I^A\la{\eta}\qquad
\la{\eta_i}\ds{\theta^A Q_A} = \sqrt{2}\ds{\theta^A i_I}\frac{\partial}{\partial\eta_I^A}\la{\eta}
\end{align}
where small $\ra{\theta_A}$ and $\rs{\theta^A}$ parameterise a linearised supersymmetry transformation. The supercharges may therefore be represented as linear operators on the superamplitudes 
\begin{equation} \label{eqn:ampsupercharges}
Q^{\dagger A}=-\sqrt{2}\sum_i\ra{i^I}\eta_{i,I}^A\qquad Q_{A}=\sqrt{2}\sum_i\rs{i_I}\frac{\partial}{\partial\eta^A_{i,I}}
\end{equation} 
or on individual legs:
\begin{equation} \label{eqn:ampspinorsupercharges}
\begin{split}
q^{\dagger A}_{i, I}=-\eta_{i, I}^A&\qquad q^{ I}_{i, A}=-\frac{\partial}{\partial\eta_{i,I}^A}.
\end{split}
\end{equation}

Supersymmetry transformations of both types act simply on these coherent states:
\begin{equation} \label{eqn:susyxforms}
%\begin{split}
\la{\eta} e^{i\xi^{\dagger I}_{A} q_{i,I}^{\dagger A}} =  e^{-i\xi^{\dagger I}_{A} \eta_I^{A}}\la{\eta}, \qquad \la{\eta} e^{-i\xi^A_{I} q_{i,A}^{I}} = \la{\eta + i\xi}.
\end{equation}
\noindent 
Here, $\xi_I^{A}=\ds{\theta^A i_I}$ and $\xi^{\dagger I}_{A}=\da{\theta_A i^I}$ parameterise the supersymmetry transformation projected onto the spinors of leg $i$ of the appropriate chirality. The action of the supercharges encoded in (\ref{eqn:susyxforms}) give the supersymmetric Ward identities relating the components.

As established in Appendix \ref{Review}, massless limits are most naturally taken in the helicity basis for the massive little group (in which momentum is chosen as the quantisation axis). We will adopt the convention that this frame is chosen unless specified otherwise, so that little group indices always denote helicity by default.

By construction, the massive on-shell superfields do not depend upon a preferred frame of reference. However, as a result, the difference between massless and massive representations of the supersymmetry algebra is firmly ingrained in the formalism, as the massive (non-BPS) states non-trivially represent a larger algebra. In the massless limit, following the rules established in (\ref{Review}), the form of the spacetime supercharges (\ref{eqn:ampsupercharges}) requires that the massive Grassmann variables are mapped onto the massless ones as $\eta_{i,-}\rightarrow\eta_i$. Here $\eta_i$ is the massless Grassmann variable used to construct the massless on shell superfields (as in e.g. \cite{Elvang:2011fx}). For reference, massless coherent states are defined here as \begin{align}
\la{\eta_i}&=\la{\Omega}e^{q_{i,A}\eta_{i}^A}\\
\la{\eta^\dagger_i}&=\int d^{\mathcal{N}}\eta e^{\eta^A\eta^\dagger_A}\la{\eta}.
\end{align}
Massless analogues of the previous formulae may be obtained similarly. 

However, for non-BPS states, the massless limit of the spacetime supercharges (\ref{eqn:ampsupercharges}) reduces the number of supercharges represented on the multiplet in half, leaving the definitions of the spinor-stripped supercharges $q_{i,+}^{\dagger A}$ and $q_{i,A}^{+}$ ambiguous. As a consequence, expressions obtained upon taking the massless limit directly on coherent states or their matrix entries will involve a residual Grassmann variables denoted here as $\eta_{i,+}\rightarrow \hat{\eta}_i$. This does not represent the action of a supercharge, but does delineate a division of the massive superfield into separate massless representations.

\subsection{Superfields}\label{SuperFields}

The coherent state construction generically gives component fields in reducible representations of the little group and $R$-symmetries, which need to be disentangled to locate the field content. The structure of these vary significantly with the number of supersymmetries. 

We consider first the simple case of $\mathcal{N} = 1$. The states in the multiplet are generated by acting $q^I$ on the Clifford vacuum and then decomposing the resulting little group tensors into irreducible representations, which will be further constrained by the needed fermionic antisymmetry of the supercharges. Choosing the Clifford vacuum to be a scalar $\phi=\la{\Omega}$, the resulting states are then $\chi^I=-\la{\Omega}q^I$ and $\tilde{\phi}^{IJ}=-\la{\Omega}q^Iq^J=\epsilon^{IJ}\la{\Omega}\frac{-1}{2}q_Kq^K$. Because the tensors are antisymmetric, the state $\tilde{\phi}^{IJ}=-\epsilon^{IJ}\tilde{\phi}$ is decomposed into a single scalar degree of freedom. The states of the chiral supermultiplet may therefore be arranged into coherent state
\begin{equation}
\Phi = \phi + \eta_I \chi^I - \half \eta_I \eta^I \tilde{\phi}
\label{chiralsup}
\end{equation}
\noindent All states in the multiplet must have identical internal quantum numbers (except for possible $U(1)_R$ charges). If the multiplet is self-conjugate, then the fermion is Majorana and the scalars are permitted to have opposite $R$-charges. Otherwise (as is necessary if the field is in a complex representation, like a quark in superQCD), an anti-superfield is required with conjugate internal quantum numbers which may be constructed similarly.

Component fields can be extracted from the full superfield via Grassmann derivations. In this simple case we have the mapping
\begin{equation}
\phi = \Phi \big|_{\eta_I = 0}, \qquad \chi^I = \frac{\partial}{\partial \eta_I} \Phi \big|_{\eta_I = 0}, \qquad \tl{\phi} = \half \frac{\partial}{\partial \eta_I}\frac{\partial}{\partial \eta^I} \Phi \big|_{\eta_I = 0}
\end{equation}

\noindent which generalizes straightforwardly to other theories. By the equivalence of Grassmann differentiation and integration, the derivatives may be replaced by integration. The Grassmann differential operators above can be used to extract component amplitudes in the usual way as for massless superamplitudes.

In the massless limit, the superfield decomposes into components that may be described by opposite helicities: 
\begin{align}
\Phi \rightarrow\Phi^+\hat{\eta}+\Phi^-
\end{align}
with
\begin{align}
\Phi^+=\chi^++&\eta\tilde{\phi},\qquad\Phi^-=\phi+\eta\chi^-. \label{MasslessChiral}
\end{align}
The limit is taken by simply replacing $\eta_-\rightarrow\eta$ and $\eta_+\rightarrow\hat{\eta}$ in (\ref{chiralsup}). Here $\eta$ is the Grassmann number that would represent the supercharge that acts non-trivially on the massless multiplet, while $\hat{\eta}$ is the variable corresponding to the trivially-acting component. 

Similarly to the extraction of component states above, each resulting massless superfield may be extracted by either setting $\hat{\eta}=0$ ($\Phi^-$ in this example) or differentiating with $\frac{\partial}{\partial\hat{\eta}}$ ($-\Phi^+$ here). This likewise allows for the extraction of massless superamplitudes from limits of massive ones.

We can next construct a vector superfield by starting with a fermionic Clifford vacuum. Because the two spin components of the fermion belong to different supermultiplets (that is, the vector multiplet does not contain its $CPT$-conjugate states), a little group covariant representation necessitates that two multiplets be combined to create an on-shell superfield that itself transforms in a non-trivial representation of the $SU(2)$ little group, where each multiplet contains states of opposite spin projections. Here, this amounts to combining two Clifford vacua into an $SU(2)$ fundamental representation to describe the two polarisation states of a fermion's degrees of freedom. The superfield is
\begin{equation}
\mathcal{W}^I = \lambda^I + \eta^I H + \eta_J W^{(IJ)} - \half \eta_J \eta^J \tilde{\lambda}^I,
\label{vectormult}
\end{equation}

\noindent where the components have already been decomposed to give the spin-$1/2$ fermion highest-level state $\tilde{\lambda}$, while we have both a real scalar $H$ and a massive vector $W^{(IJ)}$ at the first level. We can extract the different irreducible representations of the little group via $\half \frac{\partial}{\partial \eta^I} \mathcal{W}^I = H$, and $\half \left(\frac{\partial}{\partial \eta_J} \mathcal{W}^I + \frac{\partial}{\partial \eta_I} \mathcal{W}^J\right) = W^{(IJ)}$. 

Taking the massless limit again, the massive vector supermultiplet decomposes into the two helicity components of a massless vector superfield and those of a massless chiral superfield as 
\begin{alignat}{2}
\mathcal{W}^+&\rightarrow G^+\hat{\eta}+\Phi^+\qquad &&\mathcal{W}^-\rightarrow G^-+\Phi^-\hat{\eta},\nonumber\\
G^+&=g^++\eta\tilde{\lambda}^+\qquad &&G^-=\lambda^-+\eta g^-,\label{MasslessLimVector}\\
\Phi^+&=\lambda^++\eta\Big(\frac{1}{\sqrt{2}}W^L+H\Big)\qquad &&\Phi^-=\Big(\frac{1}{\sqrt{2}}W^L-H\Big)+\eta\tilde{\lambda}^-\nonumber
\end{alignat}
The longitudinally polarised vector, $W^L=\sqrt{2}W^{(+-)}$, combines with the scalar $H$ to give the two real scalar degrees of freedom in the massless chiral superfields. 

For $\mathcal{N} = 2$ without a central extension, we essentially just have two copies of the $\mathcal{N} = 1$ superalgebra. There is only one supermultiplet to construct here, namely that which starts with a scalar Clifford vacuum, as any other choice takes us into supergravity. The other familiar $\mathcal{N} = 2$ supermultiplets are short multiplets. Expanding the coherent state and keeping the $R$-indices gives the superfield 
\begin{equation}
\Omega =\phi+\eta^A_I\psi^I_A-\half \eta^A_I \eta^B_J (\epsilon^{IJ} \phi_{(AB)} + \epsilon_{AB} W^{(IJ)})+\frac{1}{3}\eta_I^B\eta_{JB}\eta^{JA}\tilde{\psi}_A^I +\eta_1^1\eta_1^2\eta_2^1\eta_2^2\tilde{\phi}
\label{LongMultNtwo}.
\end{equation}
Each term has been decomposed into irreducible little group and $R$ components (remembering that $\epsilon_{AB}$ may be used to raise and lower the $SU(2)$ $R$-indices). The Grassmann order $3$ and $4$ terms respectively represent a pair of chiral fermions related by $R$-symmetry and a scalar. The fermion $\tilde{\psi}_A^I$ may be extracted by the action of the Grassmann derivatives for each of the Grassmann variables except $\eta_I^A$. Fermion statistics of the Grassmann generators implies that the Grassmann order $2$ terms must be symmetric in either little group or $R$ indices, hence the little group triplet vector and $R$-triplet scalars.
The scalars $\phi_{(AB)}$ may be extracted by the action of $\frac{1}{2}\frac{\partial}{\partial\eta_I^A}\frac{\partial}{\partial\eta^{IB}}$, while the vectors $W^{(IJ)}$ are extracted by $\frac{1}{2}\frac{\partial}{\partial\eta_I^A}\frac{\partial}{\partial\eta_{JA}}$. This superfield will be discussed further in \cite{Herderschee:2019dmc} as a short multiplet in $\mathcal{N}=4$ super-Yang-Mills.

Of course, higher-spin representations - either fundamental supergravity multiplets or composite superfields - may be constructed using the same methods. For example, a general massive $\mathcal{N}=1$ superfield $S$ of spin $s$ has the form
\begin{equation}
S^{(I_1\dots I_{2s})} = \phi^{(I_1\dots I_{2s})} + \eta^{(I_1} \psi^{I_2\dots I_{2s})} + \eta_{J} \Psi^{(J I_1\dots I_{2s})} - \frac{1}{2} \eta_J \eta^J \tilde{\phi}^{(I_1\dots I_{2s})},\label{HigherSpinMassive}
\end{equation}
where $\psi$, $\phi$, $\tilde{\phi}$ and $\Psi$ are its component states in order of increasing spin. In the massless limit, this decomposes into pairs of separate superfields each containing either one $\phi$ state (with helicity between $-s$ and $s$) or one $\tilde{\phi}$ state (the superfield having helicity between $-s+\frac{1}{2}$ and $s+\frac{1}{2}$). For reference, a massless higher spin superfield with Clifford vacuum of helicity $h$ is 
\begin{align}
\Sigma^{h}=\varphi^h+\eta\xi^{h-\frac{1}{2}},\label{HigherSpinMassless}
\end{align}
where $\varphi$ and $\xi$ are the component states.

\section{Constructing and Constraining On-Shell Superamplitudes} \label{sec:gen3amps} \label{sec:examples}

We wish to write down superamplitudes which package together the scattering data for full representations of the super-Poincar\'{e} algebra and allow for amplitudes of component states to be projected out in a simple manner. In this form the SUSY Ward identities will be simply represented. We first discuss general features of superamplitudes in \ref{sec:generalamps} with a focus on three legs, and then lay out useful strategies for building them in \ref{Strategies}, with a focus on $\mathcal{N} = 1$. In \ref{sec:Parity} we discuss the imposition of parity symmetry at the level of the superamplitude.  We assume in this section the absence of central charges.

\subsection{SUSY Invariants and the \texorpdfstring{$\eta, \eta^\dagger$}{eta, eta-conjugate}  Bases} \label{sec:generalamps}

Invariance under supersymmetry implies that each $n$-leg superamplitude, $\mathcal{A}_n$, must be annihilated by the supercharges. In the $\eta$ basis defined in Section \ref{sec:multiplets}, the multiplicative action of $Q^\dagger$ implies that $Q^\dagger \mathcal{A}_n=0$ is solved if and only if $\mathcal{A}_n$ is proportional to the delta function
\begin{equation}
\delta^{(2\mathcal{N})}(Q^\dagger) = \prod\limits_{A=1}^{\mathcal{N}} \left( \sum\limits_{i<j}^n \da{i^I j^J} \eta_{iI}^A \eta_{jJ}^A + \half \sum\limits_{i}^n m_i \eta_{iI}^A \eta_{i}^{IA} \right).
\end{equation}
A straightforward calculation using momentum conservation shows that this delta function is also invariant under supersymmetry transformations by $Q_{A\alpha}$. However, as these transformations are not multiplicative, this does not exhaust the constraints from $Q$ transformations.

If we had instead put all of our external states in the $\eta^\dagger$ representation, the $Q$ supercharges would act multiplicatively and the Ward identity $Q\mathcal{A}_n=0$ would instead imply that the amplitude is proportional to the delta function 
\begin{equation}
\delta^{(2\mathcal{N})}(Q) = \prod\limits_{A=1}^{\mathcal{N}} \left( \sum\limits_{i<j}^n \ds{i_I j_J} \eta_{iA}^{\dagger I} \eta_{jA}^{\dagger J} + \half \sum\limits_{i}^n m_i \eta_{iA}^{\dagger I} \eta_{iIA}^{\dagger} \right).
\end{equation}
The Fourier transform of this delta function, $\widetilde{\delta^{(2\mathcal{N})}(Q)} = \int \prod\limits_{i=1}^n \text{d}^{2\mathcal{N}}\eta^\dagger_i e^{- \eta^A_{iI}\eta_{iA}^{\dagger I}} \delta^{(2\mathcal{N})}(Q)$, is also a supersymmetric invariant in the $\eta$ basis, as can be seen by commuting $Q, Q^\dagger$ through the exponential. For amplitudes with massive particles, including three-leg amplitudes, this Fourier transformed delta function is always of degree at least as large as $\delta^{(2\mathcal{N})}(Q^\dagger)$. 

Exceptions do exist in situations involving three-particle superamplitudes between BPS states in theories with extended supersymmetry. This will be elaborated upon further in \cite{Herderschee:2019dmc}, but we will merely comment here that, in these cases, some subset of the supercharges degenerate. The supersymmetric invariant in will instead be the product of all of the distinct supercharges. This is similar to the case of three massless particles, where, for example, special kinematics can imply that if the square brackets are nonvanishing, then $\da{Q^{\dagger A} Q^{\dagger A}} = 0$. The supersymmetric invariant must instead be taken as $\prod\limits_{A=1}^{\mathcal{N}} \frac{\ds{23}}{\da{q1}} \da{qQ^{\dagger A}} =  \prod\limits_{A=1}^{\mathcal{N}} \left( \ds{23} \eta_1^A + \ds{31} \eta_2^A + \ds{12} \eta_3^A\right)$, with $\ra{q}$ a reference spinor satisfying $\da{qi}\neq 0$ for all $\ra{i}$, which matches $\widetilde{\delta^{(2\mathcal{N})}(Q)}$. 

The existence of the different $\eta$ or $\eta^\dagger$ bases for the same superamplitude yields a restriction on its maximum Grassmann degree from knowledge that the delta functions are the lowest Grassmann degree invariants. 
This restriction is especially important for the construction and classification of three-leg superamplitudes. For any number of massive external particles, we can always write a three-leg superamplitude in either basis as 
\begin{equation}
\begin{split}
\left.\mathcal{A}_3\right|_\eta = \delta^{(2\mathcal{N})}({Q^\dagger}){F}({\eta}_{I}) \\
\left.\mathcal{A}_3\right|_{\eta^\dagger} = \delta^{(2\mathcal{N})}(Q)\bar F(\eta^{\dagger I})
\end{split}
\end{equation}

\noindent where $F,\bar F$ are so far undetermined and are also functions of momentum spinors. Na\"{i}vely, these functions could have maximum Grassmann degree $\mathcal{N}(M + 3)$, where $M$ is the number of massive legs, since this is the number of independent Grassmann variables we have.

However, from above we know that the Grassmann Fourier transform relates fields in the $\eta$ basis to those in the $\eta^\dagger$ basis and thus such a transform of all legs relates the superamplitude in the two bases. That is, $\widetilde{\left.\mathcal{A}_3\right|_\eta} = \left.\mathcal{A}_3\right|_{\eta^\dagger}$. The Grassmann Fourier transform roughly returns the set complement of the $\eta^\dagger$s in the original expression from the total number of $\eta$s (a full discussion of the Grassmann Fourier transform may be found in Appendix \ref{Grassmann}). So we end up with $\left[\widetilde{\left.\mathcal{A}_3\right|_\eta}\right]_{\eta^\dagger} = \mathcal{N}(M + 3) - 2\mathcal{N} - \left[F\right]_{\eta} = \mathcal{N}(M + 1) - \left[F\right]_{\eta}$, denoting by $\left[X\right]_{\eta}$ the Grassmann degree in $\eta$ of some polynomial $X$. However, the Grassmann degree of $\left.\mathcal{A}_3\right|_{\eta^\dagger}$ is at least $2\mathcal{N}$, because this is the minimal Grassmann degree for the SUSY invariant to which it must be proportional. Hence we have the inequality 
\begin{equation}
\left[F\right]_{\eta} \leq \mathcal{N}(M - 1)
\label{Gdeg}
\end{equation}

\noindent Of course, the same reasoning holds with $F$ replaced with $\bar F$. \footnote{As remarked previously, the situation is modified in the case of three massless particles because there is a SUSY invariant with Grassmann degree $\mathcal{N}$. In this case, SUSY directly implies that the only possible Grassmann structures are $\delta^{(2\mathcal{N})}(Q^\dagger)$ and $\widetilde{\delta^{(2\mathcal{N})}(Q)}$.
The case in which the particles are BPS is also exceptional and will be explained in \cite{Herderschee:2019dmc}.} 

\indent This simplifies our task of constructing general three-leg superamplitudes as we need only understand the structure of appropriately invariant functions of Grassmann degree $2\mathcal{N}$ at most.

\subsection{Strategies for Enumerating Amplitudes without Central Charges} \label{Strategies}

The main goal will be to construct three-leg superamplitudes in all simple supersymmetric theories with spins $\leq 1$. We presently discuss the procedure in brief and outline a number of simplifications. 

Now that we only have a small number of Grassmann orders to worry about at most, our task will be to construct the function $F$ which multiplies the SUSY invariant delta function for various theories. This function is constrained by the little group covariance of the amplitude, which is set by the external legs as in the $\mathcal{N}=0$ case. Supersymmetrically, it is constrained by the Ward identities, since half of the supercharges act derivatively. An important benefit of our representation of the supercharges (\ref{eqn:ampsupercharges}) is that they are of uniform degree in $\eta$ and consequently these constraints do not mix up different Grassmann orders. This simplifies the procedure so that we may construct the amplitude order by order in $\eta$. 

At each order, the $F$ factor consists of a sum of monomials in Grassmann variables. As the delta function is little group invariant, each of these terms must carry the little group representations of the superfield legs. The Grassmann variables themselves transform in non-trivial little group representations, so must be combined with coefficients built out of spinors in such a way as to give the necessary representation of the superamplitude. The possible combinations of spinors that satisfy this then correspond to the possible terms that are permitted by supersymmetry and Lorentz invariance. For example, a superamplitude with three massive spin-$1/2$ legs will have an $F$ factor with a single spin index for each leg ($F^{I_1J_2K_3}$). An example of a candidate term with Grassmann degree $1$ is then $c^{I_1J_2K_3M_1}\eta_{1M_1}$, where the Grassmann variable from leg $1$ contains a little group index for that leg, while the coefficient's tensor structure is then determined by that of $F^{IJK}$ and $\eta_{1M}$. 

Each Grassmann variable carries either a fundamental $SU(2)$ index for a massive leg or a helicity weight of magnitude $1/2$ for a massless leg. The rank and helicity weight of the representations of the possible coefficients are determined by the possible combinations of Grassmann monomials with the required little group structure. We define the `total little group weight' $\mathbf{h}$ of a superamplitude to be 
\begin{align}
\mathbf{h}\,\,=\sum_{\text{massive legs}}2s_i\,\,+\sum_{\text{massless legs}}2|h_i|,
\end{align}
for spin $s_i$ (helicity $h_i$) of the massive (massless) leg $i$.

Each coefficient of the Grassmann monomials must involve an even number of contracted spinors (as the superamplitude is a Lorentz scalar). This implies that terms with an even number of Grassmann variables cannot arise if $\mathbf{h}$ is odd for the amplitude. Likewise, if the amplitude as a little group tensor has even $\mathbf{h}$, only even Grassmann degree terms are consistent.

The possible tensor coefficients of the Grassmann monomials may be constructed similarly to the way in which $S$-matrix amplitudes are constructed in \cite{Arkani-Hamed:2017jhn}. The coefficients like $c^{I_1J_2K_3M_1}$ above may be expanded in a tensor basis spanned by a massive spinor of our choice for each of the required little group indices. We are then left to construct, for each monomial, an $SL(2,\mathbb{C})$ Lorentz tensor with the correct little group weight for the massless legs and massless Grassmann variables, which we may do by identifying a tensor basis and enumerating the possibilities as done in \cite{Arkani-Hamed:2017jhn}. 

A similar procedure to that used in \cite{Elvang:2009wd} may be used to determine $F$. As $F$ is ambiguous up to the addition of terms $\propto Q^\dagger$ (as these are annihilated by $\delta(Q^\dagger)$), arbitrary linear combinations of this supercharge may be added to simplify the superamplitude. The two components of each supercharge can be used to eliminate two Grassmann variables entirely from $F$ (for example, the two little group components of a particular massive leg). We may then apply the supersymmetry constraint $QF = 0$ to relate the spinor coefficients of different Grassmann monomials to each other.

An exceptional feature appears in the special case of a three-leg amplitude for two massive, equal-mass particles and one massless particle. In this special kinematic configuration, one finds that there is an additional object that can carry the little group weight of the massless particle. Following \cite{Arkani-Hamed:2017jhn}, this is
\begin{equation}
x \equiv \frac{1}{m} \frac{\ls{q}p_2 \ra{3}}{\ds{q3}}, \label{eqn:xdef}
\end{equation}
where $3$ is the massless leg, $m$ is the mass of legs $1$ and $2$, and $\rs{q}$ is an arbitrary reference spinor defined so that $\ds{q3}\neq 0$. In this unique case, the special kinematics of the legs implies that $p_1\cdot p_3=-\la{3}p_1\rs{3}=0$ and so $p_1\rs{3}\propto \ra{3}$. The constant of proportionality is $x$, which, as a $SL(2,\mathbb{C})$ scalar, nevertheless carries helicity weight $1$ of leg $3$. In no other kinematic configuration of massive legs in a $3$-particle amplitude does such an alignment of massless spinors occur in which the relative orientation is described by a single scalar. This is the reason that (\ref{eqn:xdef}) is independent of the reference spinor, despite its necessary appearance in inverting $\frac{1}{m}p_1\rs{3}=x\ra{3}$, and also the reason that such a helicity-weight carrying scalar object doesn't exist in other kinematic configurations. 

With this general method established, we turn to the construction of elementary amplitudes in simple SUSY theories, after first digressing to discuss parity.

\subsection{Parity}\label{sec:Parity}
While not obligatory, many theories exhibit parity ($P$) symmetry. We here explain how this acts upon on-shell superfields, from which relations between superamplitudes in a parity-invariant theory may be deduced. Details about the construction and spin quantisation of spinors can be found in Appendix C of \cite{Dreiner:2008tw}.

As for general chiral spinors, parity acts on the super-Poincare group as \cite{Weinberg:2000cr}
\begin{align}\label{parity}
P P_\mu P^\dagger = \mathcal{P}^\nu_\mu P_\nu\qquad P Q_\alpha P^\dagger = iQ^{\dagger\dot{\alpha}}\qquad P Q^\dagger_{\dot{\alpha}} P^\dagger = -iQ^{\alpha}.
\end{align}
where $\mathcal{P}^\nu_\mu=\text{diag}(1,-1,-1,-1)$.

The action of parity on the coherent states may be determined by its action on the Clifford vacuum and on the spinor-stripped supercharges $q$, defined in (\ref{supercharges}). It is important to remember here that these have been implicitly defined with restriction to a particular momentum eigenspace. The operators $q_i$ and $q^\dagger_i$, through their particle labels, implicitly also carry momentum labels. Under the action of parity, they are mapped to their representations on different momentum eigenspaces.

For massless legs, noting that 
\begin{align}\label{SpinorP}
\ra{\mathcal{P}p}=-e^{i\varphi}\rs{p}\qquad \rs{\mathcal{P}p}=e^{-i\varphi}\ra{p}\nonumber\\
\la{\mathcal{P}p}=e^{i\varphi}\ls{p}\qquad \ls{\mathcal{P}p}=-e^{-i\varphi}\la{p}
\end{align}
for a phase $\varphi$, the action of $P$ on the supercharges $q_i$ and $q^\dagger_i$ is derived from (\ref{parity}) to be
\begin{align}\label{SuperPmassless}
P q_iP^\dagger=-ie^{i\varphi}q^\dagger_{\mathcal{P}i}\qquad P q_i^\dagger P^\dagger=ie^{-i\varphi}q_{\mathcal{P}i}.
\end{align}
Here, $\mathcal{P}i$ denotes leg $i$ with inverted $3$-momentum. Note that helicity spinors are defined up to a convention-dependent, arbitrary overall phase, which must be implicitly made in the definition of the spinor-stripped supercharge. This effectively determines an arbitrary phase multiplying the (complex) Grassmann variables $\eta^A$ in the coherent states, which, as will be shown below, can be defined to absorb these factors in the parity-conjugate superfield.

The existence and action of $P$ is a model-dependent property. Depending upon the theory, supermultiplets may be self-conjugate or mapped to distinct supermultiplets. Massless spinning particles must be mapped to states with opposite helicity, which are usually part of a distinct supermultiplet. However, because of (\ref{parity}), massless scalars  and massive spinning particles, at least when selected as Clifford vacua, must also be mapped to states of distinct weight (in the same or possibly different multiplets) for consistency with SUSY. For theories with this property, the action of $P$ on a massless coherent state may be determined as follows. Taking for example the left-handed chiral multiplet $\Phi^-$ in (\ref{MasslessChiral}) and explicitly labeling its $3$-momentum $\vec{p}$,
\begin{align}
\Phi^-_{\vec{p}}P^\dagger &=\la{\phi_{\vec{p}}}P^\dagger Pe^{q_{\vec{p}}\eta_{\vec{p}}}P^\dagger\nonumber\\
&=\zeta_{\phi}\la{\tilde{\phi}_{-\vec{p}}}e^{\eta^\dagger_{-\vec{p}}q^\dagger_{-\vec{p}}}=\zeta_{\phi}\widetilde{\Phi^+_{-\vec{p}}},
\end{align}
calling the Grassmann variable of the parity conjugate coherent state $\eta^\dagger_{-\vec{p}}=ie^{i\varphi}\eta_{\vec{p}}$, absorbing the phase from the transformation of the supercharge. Here, the $\widetilde{\Phi^+}$ denotes Grassmann Fourier transform of the chiral superfield $\Phi^+$ in (\ref{MasslessChiral}) (which, in general, need not have any other necessary relationship with $\Phi^-$). Finally, $\zeta_{\phi}$ is an intrinsic parity assigned to the scalars (note that the Clifford vacuum is not a parity eigenstate).

Similarly, 
\begin{align}
\Phi^+_{\vec{p}}P^\dagger = \zeta_{\lambda^+}\widetilde{\Phi^-_{-\vec{p}}}\\
G^+_{\vec{p}}P^\dagger = \zeta_{g^+}\widetilde{G^-_{-\vec{p}}}\label{ParityG+}\\ 
G^-_{\vec{p}}P^\dagger = \zeta_{\chi^-}\widetilde{G^+_{-\vec{p}}}
\end{align}
where the Clifford vacuum for $\Phi^+$ maps under parity as $\la{\lambda^+_{\vec{p}}}P^\dagger=\zeta_{\lambda^+}\la{\lambda^-_{-\vec{p}}}$ (and analogously for the other coherent states). The factors of $\zeta_X$ are possible phases associated with intrinsic parity of the Clifford vacuum. For example, in SUSY QED, the action of parity on the photon's multiplet would introduce a factor of $\zeta_{g^+}=-1$ in (\ref{ParityG+}) because of the intrinsic parity of the photon.

For massive legs, the null vectors that constitute the little group decomposition of the massive momenta transform in the same way as (\ref{SpinorP}) under $3$-momentum inversion:
\begin{align}\label{SpinorPmassive}
\rs{\mathcal{P}p^I}=\ra{p^I}&\qquad 
\ra{\mathcal{P}p^I}=\rs{p^I}\nonumber\\
\ls{\mathcal{P}p^I}=-\la{p^I}&\qquad 
\la{\mathcal{P}p^I}=-\ls{p^I}.
\end{align}
Helicity reverses under parity because, while spin is invariant, the quantisation axis (defined as the $3$-momentum) reverses. The massive little group components are expressed with respect to some external quantisation axis, rather than the $3$-momentum, so should not change under parity. This is the reason that the phases that accompanied the transformation of the massless helicity spinors (and subsequently the supercharges) do not arise here. The massive supercharges therefore transform as
\begin{align}\label{SuperPmassive}
P q_{i}^IP^\dagger=iq^{\dagger I}_{\mathcal{P}i}\qquad P q_{i,I}^\dagger P^\dagger=iq_{\mathcal{P}i,I}.
\end{align}

\noindent Calling $\eta^\dagger_{-\vec{p},I}=i\eta_{\vec{p},I}$, the action of $P$ on a massive chiral superfield is 
\begin{align}
\Phi_{\vec{p}}P^\dagger=\zeta_\phi\la{\tilde{\phi}'_{-\vec{p}}}e^{q^{\dagger I}_{-\vec{p}}\eta^\dagger_{-\vec{p}I}}=\zeta_\phi\widetilde{\Phi'_{-\vec{p}}},
\end{align}
where, depending upon other quantum numbers, $\Phi'$ may or may not be equal to $\Phi$. The scalar Clifford vacuum is importantly mapped to the scalar of opposite weight in the other superfield: $\phi\rightarrow\tilde{\phi'}$. For a massive vector, the transformation is similar but with fermionic Clifford vacuum mapped to the other fermionic degree of freedom in the multiplet with the same polarisation
\begin{align}
\mathcal{W}^I_{\vec{p}}P^\dagger=\zeta_{\chi^I}\widetilde{\mathcal{W}'_{-\vec{p}}}^I,
\end{align}
where again $\mathcal{W}'$ may or may not be distinct from $\mathcal{W}$.

Parity invariance of a theory implies equality of the superamplitudes of a set of particles with that of their parity conjugates. Given the results above, this may be stated as 
\begin{align}
\mathcal{A}_n(X_{p_1},X_{p_2},\ldots X_{p_n})=\left(\prod_{i=1}^n\zeta_{X_i}\int \text{d}^{2}\eta_{\mathcal{P}i} \ e^{\eta_{\mathcal{P}i,I}\eta^{\dagger I}_{\mathcal{P}i}}\right)\mathcal{A}_n(X^P_{\mathcal{P}p_1},X^P_{\mathcal{P}p_2},\ldots X^P_{\mathcal{P}p_n}),\label{ParityRelation}
\end{align}
where $X^P$ is the parity conjugate superfield of $X$ (while we have written the Fourier transforms in (\ref{ParityRelation}) in the form specific for massive coherent states, they should be reinterpreted as their massless analogues for each massless leg). In other words, to relate couplings between superamplitudes in a parity symmetric theory, any superamplitude $\mathcal{A}_n(X_{p_1},X_{p_2},\ldots X_{p_n})$ must be equal to that obtained by taking the superamplitude of the parity conjugate multiplets, Fourier transforming and then reversing the $3$-momenta using (\ref{SpinorP}), (\ref{SuperPmassless}), (\ref{SpinorPmassive}), (\ref{SuperPmassive}) and the relations between Grassmann variables $\eta^\dagger_{-\vec{p},I}=i\eta_{\vec{p},I}$ and $\eta^\dagger_{-\vec{p}}=ie^{i\varphi}\eta_{\vec{p}}$. Kinematic-dependent phases appearing in (\ref{ParityRelation}) from the use of (\ref{SpinorP}) and (\ref{SpinorPmassive}) may be dropped, representing arbitrary phases in the polarisations of the external legs.

\section{\texorpdfstring{$\mathcal{N}=1$}{N=1} Three-Particle Superamplitudes}

In this section we systematically construct the possible three-particle superamplitudes for scattering of $\mathcal{N} = 1$ chiral and vector superfields and identify the types of theories to which they would belong. We furthermore discuss the dependence of the couplings on the masses of the different legs, how they behave in different limits and how they may appear in ``tree-unitary'' theories \cite{Cornwall:1974km}. We also present some simple results for higher spin multiplets. In Appendix \ref{higherLeg}, we additionally present some well-known results for higher leg amplitudes recast in the little group invariant helicity spinor language.
 
\subsection{Three Chiral Supermultiplets}\label{3chiral}

We begin with the case of three massive chiral supermultiplets, and will then find the cases with massless chiral supermultiplets via appropriate limits. Our representation of the massive chiral superfield is given in (\ref{chiralsup}). This three-point superamplitude has the general form 
\begin{equation}
\mathcal{A}(\Phi_1,\Phi_2,\Phi_3)=\delta^{(2)}(Q^\dagger)F(\eta^{I}_{i}),
\end{equation}

\noindent where, from Section \ref{sec:generalamps}, $F(\eta^{I}_{i})$ is at most a second degree polynomial and, since it has total little group weight $\mathbf{h}=0$, contains only even orders in $\eta$. Since the Ward identities do not mix different Grassmann orders, we may construct each $i$-th order Grassmann term $F^{(i)}$ separately.

We will illustrate the general procedure by explicitly deriving the superamplitude from first principles using the method described in Section \ref{Strategies}. From little group scaling, $F^{(0)}$ is fixed to be a constant which we call $\lambda$. The second-order function can be simplified by using the supercharge conservation constraint imposed by the delta function, $Q^\dagger=0$. We can use this to eliminate any dependence on $\eta_{3I}$, which then leaves us with 
\begin{equation} \label{eqn:threechiral}
F^{(2)} = b \ds{1^I 2^J} \eta_{1I} \eta_{2J} + c \da{1^I 2^J} \eta_{1I} \eta_{2J} + d_1 \eta_{1I} \eta_1^I + d_2 \eta_{2I} \eta_2^{I}.
\end{equation}

The Ward identity $QF^{(2)}=0$ is a first-order Grassmann equation and results in two independent spinor equations $(b m_2 - 2 d_1)\rs{1^I} + c p_2 \ra{1^I}=0$ and similarly with $1 \leftrightarrow 2$. The independent constraints may be found by contracting with $\ls{1_I}$ and $\la{1_I} p_2$, which allows one to solve for $d_1$ in terms of $b$ and find $c=0$. Along with the same procedure for the other equation, this yields
\begin{equation}
F^{(2)} = b \left( \ds{1^I 2^J} \eta_{1I} \eta_{2J} + \half \left( m_2 \eta_{1I} \eta^I_1 + m_1 \eta_{2I} \eta^I_2\right) \right).
\end{equation}
The full superamplitude is then 
\begin{align}\label{GeneralWZ}
\mathcal{A}(\Phi_1,\Phi_2,\Phi_3)=\delta^{(2)}(Q^{\dagger})\left[\lambda+b \left( \ds{1^I 2^J} \eta_{1I} \eta_{2J} + \half \left( m_2 \eta_{1I} \eta^I_1 + m_1 \eta_{2I} \eta^I_2\right) \right)\right].
\end{align}

When all of the legs are identical the superamplitude can be written in the manifestly exchange symmetric form
\begin{equation}\label{eqn:wzbeforecpt}
\mathcal{A}(\Phi_1,\Phi_2,\Phi_3)=\delta^{(2)}(Q^{\dagger})\left[\lambda + \frac{b'}{3m} \left(\sum\limits_{i<j} \ds{i^{I}j^{J}}\eta_{iI}\eta_{jJ} + m \sum\limits_{i} \eta_{iI}\eta_{i}^{I}\right)\right].
\end{equation}
We have here redefined the coupling $b$ to make it dimensionless. 

There are three special cases to consider corresponding to the number of different massless legs. Firstly, the massless limit $m_1\rightarrow 0$ may be taken directly on (\ref{GeneralWZ}) to produce the most general expected superamplitudes
\begin{align} \label{PhiPMLamLam}
\mathcal{A}(\Phi^-_1,\Phi_2,\Phi_3)&=\lambda\delta^{(2)}(Q^{\dagger})\\
\mathcal{A}(\Phi^+_1,\Phi_2,\Phi_3)&=-b\delta^{(2)}(Q^{\dagger})\left(\ds{12^J}\eta_{2J}+m_2\eta_1\right).
\end{align}
These expressions are independent of whether $m_2=m_3$. We have assumed that the coupling $b$ is unaffected by the limit, which is self-consistent.

Similarly, taking the subsequent limit that $m_2\rightarrow 0$ results in the superamplitudes for two massless legs:
\begin{align}
\mathcal{A}(\Phi^-_1,\Phi^-_2,\Phi_3)&=\lambda\delta^{(2)}(Q^{\dagger})\qquad \mathcal{A}(\Phi^+_1,\Phi^+_2,\Phi_3)=b\delta^{(2)}(Q^{\dagger})\ds{12},
\end{align}
while $\mathcal{A}(\Phi^\pm_1,\Phi^\mp_2,\Phi_3)=0$. It is again being assumed that the couplings present no obstruction to this, which is clearly self-consistent.

In the high energy limit, the superamplitude (\ref{eqn:wzbeforecpt}) does not diverge with inverse powers of a mass scale because of the special $3$-particle kinematics. Note that $\ds{i^+j^+}\sim\mathcal{O}(m^2/E)$ or $\da{i^-j^-}\sim\mathcal{O}(m^2/E)$ as $m\rightarrow 0$ for some (complex) energy, depending upon the kinematic configuration that is converged to (individual spinor mass limits can be found in (\ref{0masslim})). The superamplitude converges to (at leading order in energy)
\begin{align}\label{eqn:masslessWZ}
\mathcal{A}(\Phi_1,\Phi_2,\Phi_3)&\rightarrow\mathcal{A}(\Phi^-_1,\Phi^-_2,\Phi^-_3) - \mathcal{A}(\Phi^+_1,\Phi^+_2,\Phi^+_3)\hat{\eta}_1\hat{\eta}_2\hat{\eta}_3\\
\mathcal{A}(\Phi^-_1,\Phi^-_2,\Phi^-_3)&=\lambda\delta^{(2)}(Q^{\dagger})\\
\mathcal{A}(\Phi^+_1,\Phi^+_2,\Phi^+_3)&=-b'\widetilde{\delta}^{(1)}(Q),
\end{align}
where $\Phi^\pm$ are the massless superfields in the notation of (\ref{MasslessChiral}). For the latter kinematic configuration, the delta function is
\begin{align}
\widetilde{\delta}^{(1)}(Q)=\ds{23}\eta_1+\ds{31}\eta_2+\ds{12}\eta_3,
\end{align}
which is a Grassmann order $1$ supersymmetry invariant that is the Fourier transform of $\delta^{(2)}(Q)$ in the $\eta^\dagger$ basis. In the first term of (\ref{eqn:masslessWZ}), $\ds{ij}\rightarrow 0$, while in the second, $\da{ij}\rightarrow 0$. The $(-)$ sign accompanying the second term arises because the Grassmann variables $\hat{\eta}_i$ must anticommute past the fermionic $\Phi^+_i$ states. 

Note that if the limit that all particles are sent massless at the same rate is instead taken, then (\ref{eqn:masslessWZ}) is exact, rather than merely leading. The fully massive superamplitude (\ref{eqn:wzbeforecpt}) contains helicity violating couplings that, in the high energy limit, scale as mass-dependent constants and cannot be expressed as a massless superamplitude.

This massless limit is to be expected from field theory, where the three scalar component amplitudes contained in the two surviving superamplitudes are expected to vanish in the massless limit according to the superpotential. Also of note is that the massive superamplitudes (\ref{GeneralWZ}) are totally determined by two parity conjugate sets of couplings. That there are no others is not completely obvious from a Lagrangian derivation, where the possibility of spontaneous supersymmetry breaking has to be explicitly checked for a given holomorphic superpotential. Here, constraints from unbroken supersymmetry are more directly applied. It automatically follows that candidate holomorphic superpotential terms, such as tadpoles and quartics that would naively give interactions that do not conform to the structures derived here, must induce spontaneous supersymmetry breaking.

The only remaining massless superamplitudes are those of superfields with mixed helicity. These are determined by symmetries to be (up to a coupling constant)
\begin{align}
\mathcal{A}(\Phi^+_1,\Phi^+_2,\Phi^-_3)=\delta^{(2)}(Q^\dagger)\frac{1}{\da{12}}\qquad \mathcal{A}(\Phi^+_1,\Phi^-_2,\Phi^-_3)=\widetilde{\delta}^{(1)}(Q)\frac{1}{\ds{23}}.
\end{align}
However, these superamplitudes have peculiar locality properties. While non-divergent in the real momentum limit, they are also non-zero, being unsuppressed by helicity conservation \cite{Adler:2016peh}. These are the supersymmetrisations of the helicity conserving scalar-fermion-fermion $3$-leg amplitude found in \cite{McGady:2013sga}. Consistent factorisation properties of $4$-leg amplitudes were used to rule this out. Notably, while consistent with symmetries, they do not appear in the massless or high energy limit of the massive superamplitudes.

The theory of a single chiral supermultiplet has an accidental parity symmetry. This is a model-dependent and needn't be a general property of this three particle superamplitude. However, we take the opportunity to comment that parity may be imposed as described in Section \ref{sec:Parity} to relate the two otherwise independent couplings. Ignoring the possible non-trivial intrinsic parity phases, this gives $b'=\lambda$, in agreement with the massless and massive cases. The Wess-Zumino three-leg superamplitude is then
\begin{equation}\label{eqn:wzfinal}
\mathcal{A}(\Phi_1,\Phi_2,\Phi_3)= \lambda \delta^{(2)}(Q^{\dagger})\left[1-\frac{1}{3m}\left(\sum_{i<j}\ds{i^{I}j^{J}}\eta_{iI}\eta_{jJ} + m\sum_i \eta_{iI}\eta^{I}_{i}\right)\right].
\end{equation}

It would be interesting to find an on-shell condition from which the accidental parity is derived as an outcome. One would have to study higher leg amplitudes in this theory with only a single chiral supermultiplet in order to derive this feature. In this regard, it would also be interesting to find how holomorphy of the superpotential is represented in the $S$-matrix. In the case where all particles are massless, each holomorphic composite operator in the superpotential contributes a contact interaction inducing a (super)amplitude that is holomorphic in helicity. The rest of the $S$-matrix is presumably then generated by consistent factorisation involving these. Mass mixes states of different helicities, so produces a violation of this pattern of helicities induced by the holomorphic contact interactions. The on-shell superspace significantly clarifies the pattern, the foundations of which were described in \cite{Cohen:2010mi} at the level of ``seed'' $MHV$ component amplitudes with the fewest legs.

\subsection{One Massless Vector} \label{sec:SQCD}

We next turn to the case of two chiral supermultiplets interacting with a massless vector multiplet. This includes matter interactions in supersymmetric gauge theories (like superQCD). Because of this, in this section we refer to the chiral supermultiplets as quarks and the vector fields as gluons. Specifically in superQCD, the states of the quark supermultiplets arrange into the following on-shell superfields:
\begin{align}
\mathcal{Q}&=\widetilde{Q}_L+\eta_I Q^I-\frac{1}{2}\eta_I\eta^I\widetilde{Q}_R\nonumber\\
\overline{\mathcal{Q}}&=\overline{\widetilde{Q}}_L+\eta_I \overline{Q}^I-\frac{1}{2}\eta_I\eta^I\overline{\widetilde{Q}}_R,
\end{align}
where $\mathcal{Q}$ are the quark and $\overline{\mathcal{Q}}$ are the antiquark states. The $L$ and $R$ subscripts identify each of the squarks. The arrangement of the states is to be contrasted with the field-theoretic off-shell superfields. However, while we will use the symbols $\mathcal{Q}$ and $\overline{\mathcal{Q}}$ to denote the chiral superfields in what follows, we will not be committing to identifying them with any particular theory beyond what we will find to be possible to construct.

It is easily shown using the methods of Section \ref{Strategies} that a three-leg superamplitude between two massive chiral multiplets and a massless vector multiplet is impossible unless the chiral multiplets have equal mass. This case is distinguished by the existence of $x$, which will allow for expressions with the required little group scaling to be constructed.

The $G^+$ and $G^-$ superamplitudes have total little group weights $\mathbf{h} = 2$ and $\mathbf{h} = 1$ respectively. The superamplitude for the positive helicity gluon superfield is simplest to construct as little group scaling immediately gives the unique form
\begin{equation}
\mathcal{A}(\overline{\mathcal{Q}}_1,G^{+}_2,\mathcal{Q}_3)=\delta^{(2)}(Q^\dagger)\frac{g}{x},
\end{equation}
where $x$ is defined in (\ref{eqn:xdef}) and $g$ is the coupling constant (which may have suppressed dependence upon possible internal quantum numbers of the states). For the negative helicity superamplitude, little group scaling, supersymmetry invariance and the Grassmann counting rule of Section \ref{sec:gen3amps} determine the superamplitude up to a single coupling constant $b$:
\begin{align}
\mathcal{A}(\overline{\mathcal{Q}}_1,G^{-}_2,\mathcal{Q}_3)=\delta^{(2)}(Q^\dagger)bx\left(\eta_{2}+\frac{1}{2m}\left(\ds{21^{I}} \eta_{1I}+\ds{ 23^{I}} \eta_{3I}\right)\right).
\end{align}

That the superamplitudes are determined here by a single coupling constant is a reflection of the fact that the anomalous magnetic dipole moment of matter fermions in $\mathcal{N}=1$ gauge theories is exactly zero \cite{Ferrara:1974wb}. Supersymmetry determines the fermionic coupling to the gauge bosons from the scalar coupling, which has only one possible Lorentz structure. As a consequence, the supersymmetry implies that the matter-photon interaction is entirely characterised by the electric charge monopole.

Thus far we have not actually assumed anything beyond particles $1$ and $3$ having equal mass. However, these superamplitudes are antisymmetric under exchange of the two matter fields $1\leftrightarrow 3$ (through $x$). As the superfields $\mathcal{Q}$ and $\overline{\mathcal{Q}}$ are bosonic, this implies that they must be distinct (the same argument applies to couplings of matter to massless vectors without supersymmetry as well). 

Parity may additionally be imposed. Assuming that the $\mathcal{Q}$ and $\overline{\mathcal{Q}}$ multiplets are both self-conjugate under $P$ (the minimal assumption), this implies that $b=g$. Parity invariance was an assumption used in the derivation of the Lie algebra structure of the matter couplings from consistent factorisation \cite{Schuster:2008nh,Arkani-Hamed:2017jhn}, which is unaffected by the quark masses (with massless matter, $CP$ also suffices, which justifies it for chiral gauge theories). It would be interesting to clarify the role of the discrete symmetry needed to relate the amplitudes on each side of the factorisation channel. In Yang-Mills field theory, this symmetry is accidental. In the examples below, we always find parity emerge in the high energy limit of massive amplitudes, as well as the massless limits of individual legs, in the terms that match onto sensible amplitudes of massless vectors. 

It is interesting to note that the $USp(2)$ massive $R$-symmetry of the SUSY algebra is broken in this theory because the gaugino couplings distinguish between the two squark states. The identification of the squarks as $L$ and $R$ is determined by the helicity of the gaugino that couples to them (the squarks are then oppositely charged under the residual unbroken massless $U(1)_R$). This is ultimately a reflection of the breaking of the $USp(2)_R$ by parity symmetry, which distinguishes between the two squarks. This would be restored in an $\mathcal{N}=2$ gauge theory, where the gauginos are Dirac fermions.

The coupling of higher spin multiplets to photons follows a similar pattern. The superamplitude for the case of the positive helicity massless vector is
\begin{align}
\mathcal{A}(\overline{S}^{(I_1\dots I_{2s})}_1,G_2^+,S^{(J_1\dots J_{2s})}_3)=\delta^{(2)}(Q^\dagger)T^{(I_1\dots I_{2s})(J_1\dots J_{2s})}.
\end{align}
SUSY places no further constraints upon $T^{(I_1\dots I_{2s})(J_1\dots J_{2s})}$, which can be constructed as in \cite{Arkani-Hamed:2017jhn} just as a general amplitude for a photon coupled to a massive spin $s$ state. This implies that the coupling to photons of the spin $s+\frac{1}{2}$ states in the multiplet is determined by that of the spin $s$ state. For a massive particle of spin $s$, there are $2s+1$ such multipoles representing each possible independent Lorentz structure in the coupling. Following \cite{Arkani-Hamed:2017jhn}, these are
\begin{align}
T^{(I_1\dots I_{2s})(J_1\dots J_{2s})}=\frac{1}{x}\Big(c_0&\prod_{i,j=1}^{2s}\ds{1^{(I_i)}3^{(J_j)}}+\frac{c_1}{m}x\prod_{i,j=1}^{2s-1}\ds{1^{(I_i}3^{(J_j}}\ds{1^{I_{2s})}2}\ds{3^{J_{2s})}2}\nonumber\\
&+\frac{c_2}{m^2}x^2\prod_{i,j=1}^{2s-2}\ds{1^{(I_i}3^{(J_j}}\prod_{i,j=2s-1}^{2s}\ds{1^{I_i)}2}\ds{3^{J_j)}2}+\ldots\Big),
\end{align}
for coupling constants $c_i$. The additional multipole moment for the coupling of the $s+\frac{1}{2}$ state is therefore determined here entirely from the lower multipoles by SUSY. This is the generalisation of the protection of the magnetic dipole moment for supersymmetric matter fermions to higher spin states. We will see another explicit example of this in Section \ref{2mass1less}, where the electric quadrupole moment of the massive vector within the spin-half vector superfield is determined by the lower multipoles.

\subsection{One Massive Vector}

We next consider the three-leg superamplitude of two massive chiral multiplets and a massive vector multiplet, as may occur in a Higgsed gauge theory. Repeating the procedure as in previous sections, we can reduce the amplitude to 
\begin{multline}
\mathcal{A}(\mathcal{Q}_1,\mathcal{W}^I_2,\mathcal{Q}_3) = \delta^{(2)}(Q^\dagger) \Big(\frac{d_1}{m_2}\da{2^I1^J}\eta_{1J}-\frac{m_3d_2+m_2d_1}{m_1m_2}\ds{2^I1^J}\eta_{1J}\\
\qquad\qquad\qquad\qquad\qquad\qquad\qquad\qquad+\frac{d_1}{m_2}\da{2^I3^K}\eta_{3K}+\frac{d_2}{m_2}\ds{2^I3^K}\eta_{3K}\Big)\\
 = \delta^{(2)}(Q^\dagger) \Big( - \frac{d_2}{m_3} \da{2^I 1^J} \eta_{1J} + \left(\frac{(m_1^2 - m_3^2) d_2}{m_1 m_3 m_2}-\frac{d_1}{m_1}\right) \ds{2^I 1^J} \eta_{1J}\\ 
 - d_1\eta_2^I + \frac{d_2}{m_2m_3} \ls{2^I}p_3 \ra{2^J} \eta_{2J} \Big).\label{QWQ}
\end{multline}
This leaves two undetermined couplings $d_1$ and $d_2$ after imposing supersymmetry invariance. The two forms stated are useful for taking massless limits $m_3\rightarrow 0$ and $m_2\rightarrow 0$ respectively.

Taking the vector massless differs depending on whether the chiral multiplets have equal mass. In the case $m_1 \neq m_3$, one recovers solely the three-chiral superamplitudes (\ref{PhiPMLamLam}) with $b = d_2' (m_3^2 - m_1^2)/(m_1 m_3) -d_1/m_1$ and $\lambda = d_1$ (this mass scaling has been anticipated in the definition of $d_1$, as well as the assumption that it is non-zero and finite in this limit), where $d_2'=d_2/m_2$ must be finite (and hence must be suppressed by some other mass scale). This is consistent with our finding above that there was no consistent three-leg superamplitude for a massless gluon and two unequal mass chiral multiplets. 

More interestingly, if $m_1 \rightarrow m_3$ at a rate $|m_1-m_3|\sim\mathcal{O}(m_2)$ as $m_2\rightarrow 0$, then non-zero superamplitudes involving massless vector multiplets may be recovered if $d_2$ remains a dimensionless constant. The reference spinors that appear in the factors of $x$ do so through the spinor limits in (\ref{1masslim}). This leaves the parity-symmetric terms in the superQCD amplitudes with $b = g = d_2$, as well as the three-chiral superamplitudes mentioned above. As alluded to above, parity in the vector coupling emerges in this special limit.

If we instead take the third leg massless, we find smoothly 
\begin{align}
\mathcal{A}(\mathcal{Q}_1,\mathcal{W}^I_2,\mathcal{Q}_3)&\rightarrow \mathcal{A}(\mathcal{Q}_1,\mathcal{W}^I_2,\Phi^-_3)+\mathcal{A}(\mathcal{Q}_1,\mathcal{W}^I_2,\Phi^+_3)\hat{\eta}_3\\
\mathcal{A}(\mathcal{Q}_1,\mathcal{W}^I_2,\Phi^-_3)&=\frac{d_1}{m_1 m_2}\delta^{(2)}(Q^\dagger)\da{3 2^I} \left(\ds{3 1^J} \eta_{1J} + m_1 \eta_3\right) \label{QWPminus} \\
\mathcal{A}(\mathcal{Q}_1,\mathcal{W}^I_2,\Phi^+_3)&=\frac{d_2}{m_2}\delta^{(2)}(Q^\dagger)\ds{2^I3}, \label{QWPplus}
\end{align}
which are alternatively determined purely from symmetries. These expressions hold regardless of whether $m_1=m_2$ or not. It is being assumed here that $d_1$ and $d_2$ do not vanish or diverge in this limit, which is self-consistent (they may still differ from their counterparts in (\ref{QWQ}) by terms of $\mathcal{O}(m_2)$).

Taking the further $m_1 \rightarrow 0$ limit of these superamplitudes requires $d_1 \sim m_1$ in (\ref{QWPminus}) and yields 
\begin{equation}
\mathcal{A}(\Phi^+_1,\mathcal{W}^I_2,\Phi^-_3) = \delta^{(2)}(Q^\dagger)\frac{1}{m_2}\ds{2^I1},\label{WPplusPminus}
\end{equation}
where we have omitted the coupling and provided the dependence on mass necessary to realize the final massless limit smoothly (so $d_1\sim 1/m_2$ and $d_2$ constant). The case in which the chiral multiplets have the same helicity is forbidden by symmetries, so does not appear in the limit. Taking finally $m_2 \rightarrow 0$, only the transverse polarisations interact non-trivially (see comments about the superamplitudes of mixed helicity chiral supermultiplets in Section \ref{3chiral}). It is easily verified that $\mathcal{A}(\Phi^+_1,\mathcal{W}^+_2,\Phi^-_3)\rightarrow-\mathcal{A}(\Phi^+_1,G^+_2,\Phi^-_3)\hat{\eta}_1$ and $\mathcal{A}(\Phi^+_1,\mathcal{W}^-_2,\Phi^-_3)\rightarrow\mathcal{A}(\Phi^+_1,G^-_2,\Phi^-_3)$. This is expected from the Higgs mechanism if the massive vector is coupled to massless matter.

In the high energy limit (taking all masses small simultaneously at the same rate), then it can be verified that
\begin{align}
\mathcal{A}(\mathcal{Q}_1,\mathcal{W}^+,\mathcal{Q}_3) &\rightarrow \mathcal{A}(\Phi^+,G^+,\Phi^-)\hat{\eta}_1\hat{\eta}_2+\mathcal{A}(\Phi^+,\Phi^+,\Phi^+)\hat{\eta}_1\hat{\eta}_3-\mathcal{A}(\Phi^-,G^+,\Phi^+)\hat{\eta}_2\hat{\eta}_3\nonumber\\
\mathcal{A}(\mathcal{Q}_1,\mathcal{W}^-,\mathcal{Q}_3) &\rightarrow -\mathcal{A}(\Phi^+,G^-,\Phi^-)\hat{\eta}_1+\mathcal{A}(\Phi^-,\Phi^-,\Phi^-)\hat{\eta}_2+\mathcal{A}(\Phi^-,G^-,\Phi^+)\hat{\eta}_3.
\end{align}
In the $\mathcal{W}^+$ limit, the coupling $d_2$ is the cubic coupling among chiral multiplets, while $d_1$ is the parity conjugate coupling. The couplings of the chiral multiplets to the massless vectors are dermined by linear combinations of these weighted by combinations of the masses.

The possibility of distinct couplings $d_1$ and $d_2$ allows for parity violation in the massive superamplitudes and accounts for the way in which the massive amplitudes can combine together states of different helicities that would otherwise be described as having different interactions. Despite the observation that chiral multiplets coupling to massless vectors must have the same mass, there is not inconsistency with the massive multiplets having different masses in the high energy limit.

\subsection{Two Vector Superfields}

We next turn to three-leg superamplitudes with two vector superfields and one chiral superfield. Starting with the case of one massive leg, we first look at a massive chiral superfield decaying into two massless vectors. The case where the massless decay products are instead both matter fields was addressed in Section \ref{3chiral}, while no consistent superamplitude may be constructed if the massless multiplets are chiral and vector. Only the superamplitudes with massless vector multiplets of the same helicity are non-zero, following from the rules of Section \ref{Strategies}. These are (calling $m$ the nonzero mass)
\begin{align}\label{ChiralDecay}
\mathcal{A}(G^-_1,\Phi_2,G^-_3)=\delta^{(2)}(Q^\dagger)a\da{13}\qquad \mathcal{A}(G^+_1,\Phi_2,G^+_3)=\delta^{(2)}(Q^\dagger)\frac{b}{m}\ds{31}^2.
\end{align}
These superamplitudes would arise, for example, in a theory involving the quantum field couplings $[\Phi\,\mathcal{W}_A\mathcal{W}_B]_F$, for (off-shell) chiral superfield $\Phi$ and super-Yang-Mills curvatures $\mathcal{W}_{A,B}$ for some Abelian gauge groups (in other words, a massive supersymmetric axion or dilaton-like coupling). Demanding $P$ invariance would imply that $a=b$ (if the massive chiral multiplets in each superamplitude are antiparticles, then the couplings may be related by $CP$ instead). The couplings $a$ and $b$ have the expected inverse mass dimension of an irrelevant interaction. Assuming that, as defined in (\ref{ChiralDecay}), they have no further dependence on the mass of the heavy chiral multiplet, then the massless limit may be taken while holding them fixed (if they instead scale as e.g. $\propto 1/m$, then this would obstruct the limit). This gives
\begin{align}
\mathcal{A}(G^-_1,\Phi_2^-,G^-_3)=\delta^{(2)}(Q^\dagger)a\da{13}\qquad\mathcal{A}(G^+_1,\Phi^+_2,G^+_3)=\widetilde{\delta}^{(1)}(Q)b\ds{13}.\label{VCVmassless}
\end{align}
and the other components are zero.

The superamplitudes for a massive vector multiplet decaying into massless vector and chiral fields may be found similarly. Those that are permitted by the symmetries are (up to coupling constants)
\begin{align}
\mathcal{A}(\mathcal{W}^I_1,G^+_2,\Phi^+_3) &= \delta^{(2)}(Q^\dagger)\frac{1}{m}\ds{1^I2}\ds{23}\label{WGplusPplus}\\
\mathcal{A}(\mathcal{W}^I_1,G^-_2,\Phi^-_3)&=\delta^{(2)}(Q^\dagger)\da{1^I2}\label{WGminusPminus}
\end{align}
All other helicity combinations are zero. The other allowed decay channel for a massive vector multiplet was found above in (\ref{WPplusPminus}).

The massless limits of the superamplitudes (\ref{WGplusPplus}) and (\ref{WGminusPminus}) converge to the superamplitudes (\ref{VCVmassless}). Both of these massive and massless superamplitudes may have a common origin, for example in the axionic coupling suggested above, where one of the vectors may become massive through the Higgs mechanism. As in previous cases, the coupling constants for (\ref{WGplusPplus}) and (\ref{WGminusPminus}) may be related by parity.

Finally, we note that it is not possible to find a superamplitude describing the decay of a massive vector multiplet into two massless vector multiplets, which is an expression of the Landau-Yang theorem.

Continuing to the two-massive-leg case, one may construct superamplitudes for massive chiral and vector supermultiplets with a massless vector, which are independent of whether the massive multiplets have the same mass or not:
\begin{align}
\mathcal{A}(\Phi_1, \mathcal{W}_2^I, G_3^-) &= a \delta^{(2)}(Q^\dagger) \da{2^I 3} \label{PWGminus} \\
\mathcal{A}(\Phi_1, \mathcal{W}_2^I, G_3^+) &= b \delta^{(2)}(Q^\dagger) \ds{2^I 3} \left( \ds{3 1^J} \eta_{1J} - \frac{m_1}{m_2} \ds{3 2^K} \eta_{2K} \right) \label{PWGplus}
\end{align}
\noindent Taking individual legs massless, one recovers solely those amplitudes already remarked on above. The high energy behaviour of these superamplitudes is poor, scaling inversely with some mass scale contained within the couplings $a$ and $b$.

Next, the superamplitudes for two massive vector multiplets and one massless chiral multiplet may be similarly determined. Using the definitions of the massless chiral supermultiplets in (\ref{MasslessChiral}), the usual arguments determine the three-particle superamplitudes to be
\begin{align}
\mathcal{A}(\mathcal{W}^I_1,\Phi^+_2,\mathcal{W'}^J_3) &= \delta^{(2)}(Q^\dagger) F_{1+}^{IJ} \left[\frac{1}{m_1} \ds{2 1^K} \eta_{1K} + \eta_{2}\right]\label{ChiralPlus}\\
\mathcal{A}(\mathcal{W}^I_1,\Phi^-_2,\mathcal{W'}^J_3) &= \delta^{(2)}(Q^\dagger) F_{1-}^{IJ},\label{ChiralMinus}
\end{align}
where
\begin{align}
F_{1\pm}^{IJ}= d_1^{(\pm)} \da{1^I 3^J} + d_2^{(\pm)} \ds{1^I 3^J}.
\end{align}
These are again independent of whether the massive legs have equal mass or not. Taking the massless limit of the first leg, the coefficients in the $\Phi^-$ superamplitude, $d^-_i$, should have no mass dependence in order to smoothly match onto amplitudes (\ref{WGminusPminus}) and (\ref{WPplusPminus}). For the $\Phi^+$ superamplitude, both coefficients must scale as $d_i^{(+)}\sim m_1$ to return to (\ref{WGplusPplus}) and (\ref{WPplusPminus}). The couplings in both cases must be suppressed by a higher mass scale. Taking the third leg massless instead, the expected limits are obtained only if $d_i^{(+)}\sim 1/m_3$, so altogether $d_i^{(+)}\sim m_1/m_3$ to leading order in $m_1$ and $m_3$ if the limits are to be both non-trivial and unobstructed. However, in either of these cases, the resulting superamplitudes must be suppressed by other mass scales and, in this sense, are ``effective''. In contrast, taking both legs massless simultaneously is possible without introducing new mass scales. In this respect, these superamplitudes are merely a special example of the case in which the chiral multiplet is also massive, which will be explained next. 

Finally, the all-massive superamplitude for two vectors and a chiral multiplet is determined to be
\begin{align}
\mathcal{A}(\mathcal{W}^I_1,\Phi_2,\mathcal{W}'^K_3) = \delta^{(2)}(Q^\dagger) \left( F_{(0)}^{IK} + F_{(2)}^{IK}\right),\label{WWPhiMassive}
\end{align}
where
\begin{align}
F_{(0)}^{IK} &= a \da{1^I 3^K} + a' \ds{1^I 3^K}\\
F_{(2)}^{IK} &= \left(b'\da{1^I3^K}+b\ds{1^I3^K}\right)  \left[\ds{1^M2^J}  \eta_{1M} \eta_{2J} + \frac{1}{2} \left( m_2 \eta_{1L} \eta^L_1 + m_1 \eta_{2N} \eta^N_2 \right) \right].
\end{align}

In the limit that the chiral multiplet becomes massless, the coefficients match on to those of (\ref{ChiralPlus}) and (\ref{ChiralMinus}) as $b'\rightarrow d_1^{(+)}/m_1$, $b\rightarrow d_2^{(+)}/m_1$, $a\rightarrow d_1^{(-)}$ and $a'\rightarrow d_2^{(-)}$. Making the matter massive does not really affect the structure of the interactions beyond their collection into the single superamplitude. The results from taking a single vector massless instead are similar as for (\ref{ChiralPlus}) and (\ref{ChiralMinus}) and will not be elaborated upon further.

More interesting instead are the high energy limits. The superamplitude (\ref{WWPhiMassive}) consists of two parity conjugate pairs of couplings. The couplings $a$ and $b$ represent ``effective'' couplings (like those of the axion/dilaton mentioned above or loop induced interactions in a perturbative field theory) that must be suppressed by some additional mass scale (and similarly $d_1^-$ and $d_2^+$ in the case with a massless chiral multiplet). On the other hand, $a'$ and $b'$ (or $d_1^+$ and $d_2^-$) correspond to couplings of a Higgs boson to massive vectors, where the Higgs belongs to a chiral multiplet (and is not part of the multiplet eaten by the vectors with the Goldstone boson). This happens when the quartic coupling of the scalar potential originates from the superpotential (``$F$-term'').

To illustrate how the superamplitude (\ref{WWPhiMassive}) scales in the UV limit, assume that $a'=\bar{a}/v$ and $b'=\bar{b}/v^2$ for some mass scale $v$ of order the leg masses and call constants $c_i=m_i/v$ for leg masses $m_i$. The leading terms in the limit are then 
\begin{align}
\mathcal{A}(\mathcal{W}^+_1,\Phi_2,\mathcal{W}'^+_3)&\rightarrow \mathcal{A}(\Phi^+_1,\Phi^-_2,G_3^+)\hat{\eta}_3-\mathcal{A}(G_1^+,\Phi^-_2,\Phi^+_1)\hat{\eta}_1-\mathcal{A}(\Phi^+_1,\Phi^+_2,\Phi^+_1)\hat{\eta}_2\nonumber\\
\mathcal{A}(\mathcal{W}^+_1,\Phi_2,\mathcal{W}'^-_3)&\rightarrow-\mathcal{A}(G_1^+,\Phi^+_1,\Phi^-_3)\hat{\eta}_1\hat{\eta}_2\hat{\eta}_3+\mathcal{A}(\Phi^+_1,\Phi^-_2,G_3^+)\hat{\eta}_3
\end{align}
and similarly for parity conjugate states. All terms in the first line depend upon the coupling $\bar{a}$ and each term proportional to $\hat{\eta}_i$ is accompanied by a factor of $c_i$. In the second line, the first term depends upon $\bar{b}c_1c_3$, while the second has coupling constant $\bar{a}c_3$. Again, this pattern of couplings reverses for the parity conjugate limits.

However, there are also subleading terms that vanish in the massless limit that cannot be placed into massless superamplitudes. These represent the effective Goldstone boson couplings to the Higgs.

A supersymmetrised version of the argument used in \cite{Arkani-Hamed:2017jhn} to demonstrate the Higgs mechanism may presumably be made from constructing a four-leg vector superamplitude from demanding consistent factorisation into $3$-leg superamplitudes (\ref{WWPhiMassive}) on each factorisation channel. Notably, an exceptional case occurs when the Higgs couples to a massive and massless vector boson in a three-particle superamplitude, as in (\ref{PWGminus}) and (\ref{PWGplus}), which will induce unitarity-violating superamplitudes in the high energy limit in tree-level processes.

\subsection{Massive and Massless Vector Multiplet Interactions}\label{2mass1less}

Let us begin with amplitudes with two massive vector superfields and one massless vector superfield, which has two distinct cases of interest. The first is $\mathcal{A}(\mathcal{W}^I,G,\overline{\mathcal{W}}^J)$, where the two massive vector superfields $\mathcal{W}$ and $\overline{\mathcal{W}}$ have the same mass. This arises in many examples, such as the adjoint Higgsing of a simple gauge theory by a single vacuum expectation value (vev), which does not feature any $3$-leg amplitudes entirely of massive vectors. In this case, the vectors are conjugates, which is the reason for our choice of notation, although we do not need to assume this at this point. The second is $\mathcal{A}(\mathcal{W}^I,G,\mathcal{W}'^J)$, where the two massive states are distinct and of different mass. These can occur, for example, in field theories with generalised Chern-Simons terms \cite{Andrianopoli:2004sv,Anastasopoulos:2006cz,Anastasopoulos:2008jt}, where at least one of the vectors is Abelian and has a Stuckelberg mass, whereas another of the vectors may be separately Higgsed. 

As in the superQCD case above, the positive helicity gluon superfield amplitudes are determined very simply. In these cases one finds
\begin{align}
\mathcal{A}(\mathcal{W}^I_1,G^+_2,\overline{\mathcal{W}}^J_3) &= \delta^{(2)}(Q^\dagger) \ls{1^I}^\alpha \ls{3^J}^\beta \left( \frac{g}{m x} \epsilon_{\alpha\beta} + \frac{g-h}{m^2} \rs{2}_\alpha \rs{2}_\beta \right) \label{WGWbar}\\
\mathcal{A}(\mathcal{W}^I_1,G^+_2,\mathcal{W}'^J_3) &= \delta^{(2)}(Q^\dagger) \, a \ds{1^I 2} \ds{3^J 2}\label{WGW'},
\end{align}

\noindent where, in both cases, the number of free parameters matches that in the non-supersymmetric amplitude for two massive fermions and one massless vector \cite{Arkani-Hamed:2017jhn}. As in all previous examples, we have here neglected to show that the coupling constants $g$ and $h$ may have internal quantum number structure. In the first case (\ref{WGWbar}), the combination of terms with coupling $g$ corresponds to a massive vector `minimally coupled' to the massless vector. As has been foreseen in the definition of dimensionless couplings in (\ref{WGWbar}), in the limit that $m\rightarrow 0$ or, equivalently here, at energies $\gg m$, these terms converge to their expected massless counterparts. 

The term proportional to $h$ would have the perturbative interpretation of an anomalous magnetic dipole moment for the massive vector (or electric dipole moment if it has a complex phase). This term has poor behavior in the UV limit for certain helicity configurations, which is the reason for the tree-level universality of the magnetic dipole moment $h=0$ for elementary particles \cite{Ferrara:1992yc}. Note that supersymmetry has set fixed the possible quadrupole structure of the massive vector boson amplitudes that may otherwise exist as a further independent Lorentz structure in the vector boson component amplitude \cite{Arkani-Hamed:2017jhn,Guevara:2017csg}. This derivation makes obvious the way that supersymmetry determines the vector amplitudes from their fermionic counterparts. 

Finally, (\ref{WGWbar}) appears to be symmetric under exchange of particles $1$ and $3$ ($x\mapsto -x$ under this exchange - see (\ref{eqn:xdef})). However, because the superfields are fermionic, the vector multiplets must be distinct.

In the second example (\ref{WGW'}), the coupling $a$ has mass dimension $-2$. However, unlike for the minimal coupling terms in the case above, the kinematic factors of the component superamplitudes corresponding to the $+$ helicity states (such as $A(W^+G^+W'^+)$) contain terms that merely scale as $\sim\mathcal{O}(m_i)$ in the massless limit (see equations (\ref{0masslim}) in Appendix \ref{Review} for massless limits of spinors). The amplitude must therefore diverge in the high energy $E$ limit as $E/M$ for some mass scale $M$. Correspondingly, the examples of field theories cited above that feature these amplitudes are only effective up to a UV cut-off.

We can likewise find the negative helicity superamplitude purely from little group covariance and supersymmetry. From the same arguments as in the SQCD case, the Grassmann polynomial must only contain an order-one term. 
\begin{equation}
\mathcal{A}(\mathcal{W}_1^I, G^-_2, \mathcal{W}_3^J) = \delta^{(2)}(Q^\dagger) \ls{1^I}^\alpha \ls{3^J}^\beta F_{2\alpha\beta} \left(- \frac{1}{m_1}\eta_{1K} \ds{1^K 2} + \eta_{2} \right).
\end{equation}
The tensor $F_{2\alpha\beta}$ is then determined from the little group representations of the legs. In the equal mass case, this gives a superamplitude with two free parameters:
\begin{equation}
\mathcal{A}(\mathcal{W}^I_1, G^-_2, \overline{\mathcal{W}}^J_3) = \delta^{(2)}(Q^\dagger) \ls{1^I}^\alpha \ls{3^J}^\beta \left(\frac{g'}{m} x \epsilon_{\alpha \beta} + \frac{h'}{m^2} x^2 \rs{2}_\alpha \rs{2}_\beta \right) \left(- \frac{1}{m} \eta_{1K} \ds{1^K 2} + \eta_{2} \right).\label{WGWbarMinus}
\end{equation}
Exchange (anti-)symmetry between $\mathcal{W}$ and $\overline{\mathcal{W}}$ may be manifested by adding terms proportional to $Q^\dagger$ to give
\begin{align}
\mathcal{A}(\mathcal{W}^I_1, G^-_2, \overline{\mathcal{W}}^J_3)  = \delta^{(2)}(Q^\dagger) \ls{1^I}^\alpha \ls{3^J}^\beta \left(\frac{g'}{m} x \epsilon_{\alpha \beta} +  \frac{h'}{m^2} x^2 \rs{2}_\alpha \rs{2}_\beta \right)\nonumber\\ 
\times\left(- \frac{1}{2 m} \eta_{1K} \ds{1^K 2} + \eta_{2} - \frac{1}{2 m} \eta_{3K} \ds{3^K 2}  \right).
\end{align}

If parity is a symmetry of the theory under consideration, then this relates the superamplitudes of $\mathcal{A}(\mathcal{W}^I, G^-, \overline{\mathcal{W}}^J)$ as discussed in Section \ref{sec:Parity}. Assuming that the vector multiplets are self-conjugate, this requires that $g' = g$ and $h' = h$.

For the case where $m_1 \neq m_3$, the only option which has the correct scaling is $F_{2\alpha\beta} = b (p_3 \ra{2})_\alpha (p_3 \ra{2})_\beta$.  Our amplitude in this case is 
\begin{align}
\mathcal{A}(\mathcal{W}^I_1, G^-_2, \mathcal{W}'^J_3) &= \delta^{(2)}(Q^\dagger) b' \ls{1^I}p_3 \ra{2}  \ls{3^J}p_3 \ra{2} \left(- \frac{1}{m_1} \eta_{1K} \ds{1^K 2} + \eta_{2} \right)\nonumber\\
&=\delta^{(2)}(Q^\dagger) b \da{1^I2}\da{3^J2} \left(- \frac{1}{m_1} \eta_{1K} \ds{1^K 2} + \eta_{2} \right)\label{UnequalMinus}
\end{align}

\noindent where the coupling $b$ has been redefined in the second line to absorb some factors of mass. If parity is a symmetry of this theory, then one finds $b = a m_1/m_3$.

In the massless limit, the superfields are expected to decompose as shown in (\ref{MasslessLimVector}). In anticipation of the superamplitudes of the massless components being matched onto by the massless limit of the massive superamplitude, we first determine these directly from symmetries. The constraints of complex three-particle special kinematics, little group scaling and `locality', in the sense that the three-particle amplitudes do not scale as negative powers of momentum, determine that the superamplitudes of the massless supermultiplets are (neglecting coupling constants):
\begin{align}
\mathcal{A}(G_1^+,G_2^+,G_3^-)&=\widetilde{\delta}^{(1)}(Q)\frac{\ds{12}^2}{\ds{13}\ds{23}} \label{GpGpGm}\\
\mathcal{A}(G_1^-,G_2^+,G_3^-)&=\delta^{(2)}(Q^\dagger)\frac{\da{13}^2}{\da{12}\da{23}} \label{GmGmGp}\\
\mathcal{A}(\Phi_1^-,G_2^+,\Phi_3^+)&=\widetilde{\delta}^{(1)}(Q)\frac{\ds{23}}{\ds{13}}\\
\mathcal{A}(\Phi_1^+,G_2^+,\Phi_3^-)&=\widetilde{\delta}^{(1)}(Q)\frac{\ds{21}}{\ds{31}}.
\end{align}
Other superamplitudes between other possible combinations of massless superfields are also possible, but do not arise in taking the massless limit of (\ref{WGWbar}). 

Choosing a particular helicity configuration in (\ref{WGWbar}), the massless limit may be taken using the limits presented in Appendix \ref{Review} and identified with the superamplitudes above. The limits may be calculated explicitly to be
\begin{align}
\mathcal{A}(\mathcal{W}^+_1, G^+_2, \overline{\mathcal{W}}^+_3)\rightarrow 0, \qquad\mathcal{A}(\mathcal{W}^-_1, G^+_2, \overline{\mathcal{W}}^-_3)\rightarrow \mathcal{A}(G_1^-, G_2^+, G_3^-)\\
\mathcal{A}(\mathcal{W}^+_1, G^+_2, \overline{\mathcal{W}}^-_3)\rightarrow-\mathcal{A}(G_1^+,G_2^+,G_3^-)\hat{\eta}_1+\mathcal{A}(\Phi_1^-,G_2^+,\Phi_3^+)\hat{\eta}_3,
\end{align}
and similarly for $\mathcal{A}(\mathcal{W}^-_1, G^+_2, \overline{\mathcal{W}}^+_3)$. Similar results may be shown for the limits of (\ref{WGWbarMinus}). 

This demonstrates how the supersymmetrisation of the Higgs mechanism operates by combining well-defined UV amplitudes of massless chiral and vector multiplets into single superamplitudes of massive vector multiplets in the IR.

\subsection{Self-interacting Massive Vector Supermultiplets}

A similar analysis may be performed to determine the possible structure of three-leg superamplitudes of massive vector superfields. A general expression will include several special cases, such as when the vectors have equal mass and belong to the same species, as well as the case in which there is only one type of superfield, which must be constrained so that there are no vector self-interactions.

Just as for the cases considered previously, supersymmetry implies that the amplitude has the form
\begin{align}
\mathcal{A}(\mathcal{W}_1^I, \mathcal{W}_2^J,\mathcal{W}_3^K)=\delta^{(2)}(Q^\dagger)F_1^{IJKM}\Big(\eta_{1,M}+\frac{1}{m_2}\ds{1_M2^N}\eta_{2,N}\Big).\label{AllVectorMassive}
\end{align}
This is the extent to which supersymmetry determines the amplitude. 
The next step is to determine the number of independent Lorentz structures that can appear in $F_1^{IJKM}$. Altogether, there are $6$ such independent terms (up to others related by the Schouten identity and kinematic relations):
\begin{align} \label{LG4tensor}
F_1^{IJKM}=c_1\da{1^I3^K}\ds{2^J1^M}+c_2\ds{1^I3^K}\da{2^J1^M}+c_3\ds{1^I3^K}\ds{2^J1^M}\nonumber\\+c_4\da{1^I3^K}\da{2^J1^M}+c_5\da{2^J3^K}\epsilon^{IM}+c_6\ds{2^J3^K}\epsilon^{IM}.
\end{align}

One of the independent terms in this superamplitude represents a Higgs coupling, where the Higgs has a ``$D$-term'' quartic and is part of the chiral multiplet eaten with the Goldstone boson. In the Abelian Higgs theory, this is the only structure in the superamplitude. This may be identified by extracting the component amplitude of three vectors and setting it to zero. The component amplitude is 
\begin{align} \label{wwwcomponent}
A(W_1^{I_1I_2}, W_2^{J_1J_2},W_3^{K_1K_2})=F_1^{I_1J_1K_1M}\left(\delta_M^{I_2}\da{3^{K_2}2^{J_2}}-\frac{1}{m_2}\ds{1_M2^{J_2}}\da{3^{K_2}1^{I_2}}\right),
\end{align}
where external spin indices are implicitly symmetrised over. After simplification this reduces to five independent spin structures. Demanding that these vanish implies that $c_2=c_3=c_4=0,\,c_6=0$ and $c_5=-m_2c_1$, thereby reducing the number of independent couplings to one. The corresponding term in the superamplitude is then
\begin{align}
\mathcal{A}(\mathcal{W}_1^I, \mathcal{W}_2^J,\mathcal{W}_3^K)=c_1\delta(Q^\dagger)\Big(\left(-m_2\da{2^J3^K}\epsilon^{IM}+\da{1^I3^K}\ds{2^J1^M}\right)\eta_{1M} \nonumber\\
+\left(m_1\da{1^I3^K}\epsilon^{JM}-\da{2^J3^K}\ds{1^I2^M}\right)\eta_{2M}\Big),
\end{align}
which is manifestly antisymmetric under the exchange $1\leftrightarrow 2$. This constitutes one of the six independent contributions to the superamplitude (\ref{AllVectorMassive}) and is itself the three-particle superamplitude for the Abelian Higgs theory.

This contains component amplitudes of the form that would be expected in Abelian Higgs theories. For example, calling $H_i$ the scalar components of the supermultiplets, then
\begin{align}
A_3(W_1^{I_1I_2},H_2,W_3^{K_1K_2})&=\frac{\partial}{\partial\eta_{1I_2}}\Big(\frac{1}{2}\epsilon_{J_1J_2}\frac{\partial}{\partial\eta_{2J_2}}\Big)\frac{\partial}{\partial\eta_{3K_2}}\mathcal{A}(\mathcal{W}_1^{I_1}, \mathcal{W}_2^{J_1},\mathcal{W}_3^{K_1})\nonumber\\
&=-c_1m_3\ds{1^I3^K}\da{1^I3^K}.
\end{align}
Completion of the identification of this with a Higgs amplitude would require that $c_1$ be inversely proportional to some mass scale and that $c_1\sim 1/m_3^2$ as $m_3\rightarrow 0$ (and likewise for the other masses, repeating this argument with the identities of particles $1,2$ and $3$ permuted). These are the component amplitudes expected in the Abelian Higgs theory and, given the assumption that there are no vector self-interactions, $\mathcal{N}=1$ supersymmetry implies that there is only a single Lorentz structure and coupling consistent with this.

The remaining five couplings each describe superamplitudes with vector boson self-interactions. The triple gauge coupling vertex of three massive vectors has been studied extensively in the past in the context of the electroweak bosons of the Standard Model. An effective Lagrangian describing the independent Lorentz structures has been given in \cite{Hagiwara:1986vm}. The superamplitude (\ref{AllVectorMassive}) represents the supersymmetrisation of this. Supersymmetry restricts the seven independent couplings of \cite{Hagiwara:1986vm} to five. The two prohibited terms are those originating from $F^3$ terms (for Yang-Mills curvatures $F$ of the vectors), just as for massless amplitudes. 

Of the five remaining structures, one can be attributed to the Yang-Mills (tree) coupling. Just as for the Higgs couplings, the expected Yang-Mills vector self-interaction term may be identified by matching the component amplitude (\ref{wwwcomponent}) to the expected expression. Doing so imposes $c_3 = c_4 = 0, \  c_6 = m_2 c_2$ and identifies the gauge coupling as $c_2 = - 2 g/ (m_1 m_3)$. The Higgs coupling $c_5 = - m_2 c_1$ remains free. This structure, in addition to the Higgs coupling above, are distinguished as having UV limits that converge to massless three particle superamplitudes at leading order.

The remaining four couplings have poor UV scaling and correspond to field theoretic operators upon which gauge invariance is not linearly realised. Two of these (that are $CP$-odd) may be identified with the generalised Chern-Simons terms mentioned earlier (or are generated at loop-level by anomalies), while the other two correspond to the remaining two types of operators that may be constructed from vector multiplets with a single derivative. Of these, one corresponds to the anomalous magnetic dipole moment in the massless limit of one leg in (\ref{WGWbar}) and (\ref{WGWbarMinus}). Its $CP$-odd counterpart, in the massless limit, provides the same Lorentz structure, but with a different phase in the coupling. The other two couplings vanish in the limit of a massless leg on-shell.

Further conditions may be used to constrain or interpret the couplings, such as requiring good UV limits and properties of higher leg amplitudes. Two simple examples are provided by demanding that this amplitude matches onto either of the two cases discussed in Section \ref{2mass1less} in the limit that $m_1\rightarrow 0$. 

For the case where we leave $m_2\neq m_3$, we demand that (\ref{AllVectorMassive}) converge to (\ref{UnequalMinus}) and (\ref{WGW'}) for each helicity configuration of the massless vector multiplet. As long as the couplings do not scale as $\sim 1/m_1$, this requires that $c_3=-a$ and $c_4=-b$, up to terms $\propto m_1$. One of the terms with couplings $c_1$ and $c_2$ vanishes (which depends upon the helicity choice for index $I$) while the other degenerates with the $c_5$ and $c_6$ terms and so cannot be independently determined. Finally, the couplings $c_5$ and $c_6$ (up to inclusion of possible contributions from $c_1$ and $c_2$ as just described) match onto the terms in  (\ref{ChiralPlus}) and (\ref{ChiralMinus}) and may be identified with the couplings $d_1$ and $d_2$. 

In the case where the two remaining masses approach equality, $m_2,m_3\rightarrow m$ as $m_1\rightarrow 0$, we can demand that the coefficients of (\ref{AllVectorMassive}) approach (\ref{WGWbar}) and (\ref{WGWbarMinus}). This determines the coefficients to be $c_1=c_2=-g'/(m_1m)$, $c_3=h/m^2$ and $c_4=h'/m^2$, while it is required that $g'=g$ in the massless amplitudes (so parity must be an accidental symmetry if $h=h'=0$). These may be easily checked using the spinor limits provided in Appendix \ref{Review}. Again, matching onto the superamplitudes with massless vectors, the remaining couplings must be $c_5=-c_6=g'/m_1$, but may additionally have extra terms that would be determined by matching onto the amplitudes with massless matter (\ref{ChiralPlus}) and (\ref{ChiralMinus}). Unlike the previous case, these limits ensure that the mass scale of the couplings is given by $m$ and $m_1$, so that, if $h=h'=0$ (as is true at tree-level in perturbative gauge theories), the amplitudes would have the good UV limits arranged by the Higgs mechanism \cite{Arkani-Hamed:2017jhn}. Note that, as expected from (\ref{MasslessLimVector}), the superamplitudes have limits $\mathcal{A}(\mathcal{W}_1^+, \mathcal{W}_2^J,\mathcal{W}_3^K)\rightarrow \mathcal{A}(G_1^+,\mathcal{W}_2^J,\mathcal{W}_3^K)\hat{\eta}_1+\mathcal{A}(\Phi_1^+,\mathcal{W}_2^J,\mathcal{W}_3^K)$ and $\mathcal{A}(\mathcal{W}_1^-, \mathcal{W}_2^J,\mathcal{W}_3^K)\rightarrow \mathcal{A}(G_1^-,\mathcal{W}_2^J,\mathcal{W}_3^K)+\mathcal{A}(\Phi_1^-,\mathcal{W}_2^J,\mathcal{W}_3^K)\hat{\eta}_1$ and involve terms that pick up the extra Grassmann variable $\hat{\eta}_1$ for the massless superfield. A similar analysis can be performed by instead $m_2\rightarrow 0$ or $m_2\rightarrow 0$ in order to find further consistency conditions on the couplings to match onto the superamplitudes in the previous sections, but we refrain from providing the results here. These are consistent with the identifications of the couplings made above - that is, $c_3$ and $c_4$ are associated with the couplings that determined the anomalous magnetic and electric dipole moments in the limits of a massless leg, while linear combinations of $c_1$, $c_2$, $c_5$, $c_6$ correspond to the tree-level (``$D$-term'') Higgs and Yang-Mills couplings, while $c_5$ and $c_6$ also contain the other non-Yang-Mills contact interactions, such as those induced from Stuckelberg axions and anomalies.

\subsection{Higher Spin Amplitudes}
While the number of possible Lorentz structures in three-particle amplitudes typically grows significantly with the spin of the interacting particles, the case of a heavy particle decaying into two massless products is especially simple. As described in \cite{Arkani-Hamed:2017jhn}, the amplitude for a spin $s$ massive particle $\bar{\phi}$ to decay into two massless particles $\varphi_1$ and $\varphi_2$ with respective helicities $h_1$ and $h_2$ is uniquely
\begin{align}
A(\varphi^{h_1}_1,\varphi^{h_2}_2,\bar{\phi}_3^{(I_1\ldots I_{2s})})=G\ds{12}^{s+h_1+h_2}\prod_{i=1}^{s+h_2-h_1}\ds{3^{(I_i}1}\prod_{j=s+h_2-h_1+1}^{2s}\ds{3^{I_j)}2},
\end{align}
where $G$ is some coupling constant of mass dimension $[G]=-(2s+h_1+h_2-1)$. The notation is intended to indicate that all of the spin indices for the massive field are symmetrised over. It is being assumed that angular momentum selection rules permit this process to exist.

The supersymmetrisation of this is just as simple. Promoting $\varphi^{h_i}$ to massless supermultiplets (\ref{HigherSpinMassless}) with Clifford vacua of helicities $h_i$ and likewise $\phi$ to the corresponding massive multiplet (\ref{HigherSpinMassive}), then the three-particle superamplitude is also fixed as 
\begin{align}\label{3PHS}
\mathcal{A}(\Sigma^{h_1}_1,\Sigma^{h_2}_2,S_3^{(I_1\ldots I_{2s})})&=\frac{1}{m_S}\delta^{(2)}(Q^\dagger)A(\varphi^{h_1}_1,\varphi_2^{h_2},\bar{\phi}_3^{(I_1\ldots I_{2s})})\nonumber\\
&=\frac{1}{\da{12}}\delta^{(2)}(Q^\dagger)A(\xi^{h_1-\frac{1}{2}}_1,\xi^{h_2-\frac{1}{2}}_2,\phi^{(I_1\ldots I_{2s})}_3),
\end{align}
where $m_S$ is the mass of the heavy multiplet.

The superamplitude for scattering of four massless particles by exchange of a massive spinning particle may be constructed analogously to the non-supersymmetric case \cite{Arkani-Hamed:2017jhn}. Supersymmetry fixes the superamplitude to have the form
\begin{align}\label{4PHS}
\mathcal{A}(\Sigma_1^{h_1},\Sigma_2^{h_2},\Sigma_3^{h_3},\Sigma_4^{h_4})=\frac{1}{\da{34}}\delta^{(2)}(Q^\dagger)A(\varphi_1^{h_1},\varphi_2^{h_2},\xi_3^{h_3-\frac{1}{2}},\xi_4^{h_4-\frac{1}{2}}),
\end{align}
where the component amplitude $A(\varphi_1^{h_1},\varphi_2^{h_2},\xi_3^{h_3-\frac{1}{2}},\xi_4^{h_4-\frac{1}{2}})$ may be constructed out of the spinning Gegenbauer polynomials corresponding to the exchange of higher spin resonances, just as for the non-supersymmetric case \cite{Arkani-Hamed:2017jhn}. 

On a massive resonance, the superamplitude respects a supersymmetric factorisation into three-particle superamplitudes. For example, in the $s$-channel, 
\begin{align}\label{SuperFact}
\mathcal{A}(\Sigma^{h_1}_1,\Sigma^{h_2}_2,\Sigma^{h_3}_3,\Sigma^{h_4}_4)\rightarrow\int d^2\eta_P\mathcal{A}_L(\Sigma_1^{h_1},\Sigma_2^{h_2},S_P^{(I_1\ldots I_{2s})})\frac{1}{(p_1+p_2)^2}\mathcal{A}_R(\overline{S}_{-P(I_1\ldots I_{2s})},\Sigma_3^{h_3},\Sigma_4^{h_4}),
\end{align}
where the intermediate superfield has Grassmann variables $\eta^I_P$ and the Grassmann integral accounts for the sum over all states in the multiplet of the intermediate resonance. It has been chosen to represent the massive multiplet as outgoing in $\mathcal{A}_L$ and incoming in the other factor. The incoming superfield is then represented as the analytic continuation of an outgoing multiplet. Crossing relations imply that this must be the antimultiplet, hence the bar and the opposite height spin indices. The component antiparticles occupy opposite levels in the superfield.

The factorisation of the superamplitude (\ref{SuperFact}) is easily demonstrated as consistent with expectations from (\ref{4PHS}). Because of the simplicity of the three-particle superamplitudes, the Grassmann integral may be trivially evaluated using $\int d^2\eta_P\delta^{(2)}(Q_L^\dagger)\delta^{(2)}(Q_R^\dagger)=m_S\delta^{(2)}(Q^\dagger)$, where $Q_L^\dagger$ and $Q_R^\dagger$ are the supercharges associated with each respective subsuperamplitude above in (\ref{SuperFact}). This requires use of the analytic continuation rules for spinors and Grassmann variables given in \cite{Herderschee:2019dmc}, which here imply that $Q^\dagger_{i,-P}=-Q^\dagger_{i,P}$ for state $i$ of momentum $P$. The two representations of the three-particle superamplitude (\ref{3PHS}) can then be substituted to confirm that (\ref{SuperFact}) is given simply by (\ref{4PHS}) with the exhibited component amplitude factorised into the component three-particle amplitudes shown in (\ref{3PHS}).

\section{Conclusion}

We have here initiated the study of the on-shell properties of supersymmetric theories by developing the on-shell superspace formalism in which states are described in a supermultiplet by their asymptotic quantum numbers - momentum, total spin and polarisation - without the need to commit to a frame of reference. This was used to construct massive supermultiplets and represent these in scattering amplitudes of supersymmetric theories, concentrating here on $\mathcal{N}=1$ theories. Purely from the foundational principles of quantum mechanics, special relativity and supersymmetry, we constructed all of the possible elementary on-shell three-point amplitudes for multiplets of spin no greater than $1$.

A more exhaustive study into the extent to which $S$-matrix postulates constrain supermultiplets and their interactions at weak coupling is warranted. Further constraints upon theories from assumptions about IR properties, such as factorisation or behaviour in the high energy limit, remain to be investigated. 

It would be interesting to more broadly catalogue theories characterised by their spectra and interactions from conditions on IR properties and see whether they conspire to imply emergent symmetries or uniqueness properties \cite{Benincasa:2007xk,McGady:2013sga,Arkani-Hamed:2017jhn}. For example, consequences of supersymmetry on emergent properties of theories constructible from soft limits were recently investigated in \cite{Elvang:2018dco}. We do not foresee difficulties in extending our analysis to scattering states of higher-spin composite superfields or including multiplets of supergravity or Kaluza-Klein modes (see recently \cite{Caron-Huot:2018ape,Chung:2018kqs,Guevara:2018wpp} for a possible application to black holes).

To progress beyond single particle representations and $3$-leg amplitudes, some guidance for systematically constructing higher order (loop and leg) amplitudes from infrared (on-shell) properties would be desirable, such as on-shell recursion. However, because the validity of massless recursion is often sensitive to the helicity of the shifted states, the effective combining of massless states of definite helicity into massive particle representations of the (super-)Poincare group poses a potential obstruction. Prospects for overcoming this are most promising in $\mathcal{N}=4$ SYM where, for massless amplitudes, a myraid of constructibility properties have been discovered. Vestiges of these may remain present on the Coulomb branch, in particular the dual (super)conformal symmetry. In \cite{Herderschee:2019dmc} we formulate a massive super-BCFW shift and prove its validity for the construction of all Coulomb branch tree superamplitudes. The constructibility of Coulomb branch superamplitudes seems to arise from a surprising `nonlocality' present in the three-particle superamplitudes. This remains an interesting avenue for future work.

\acknowledgments 

We thank Tim Cohen, Nathaniel Craig, Henriette Elvang, and Callum Jones for comments on a draft of this work, Nathaniel Craig for discussions and support during the completion of this work, and Nima Arkani-Hamed and Yu-tin Huang for discussions on \cite{Arkani-Hamed:2017jhn}. AH and SK are grateful for the support of a Worster Fellowship. This work is supported in part by the US Department of Energy under the grant DE-SC0014129.

\appendix

\section{Conventions and Useful Identities}

\subsection{Spinor Helicity for Massive Particles}\label{Review}

We here summarise helicity spinors for massive particles and its consequences, taking the opportunity to establish the conventions and notation that we adopt throughout this article and also to present useful identities. The reader is referred to \cite{elvang2015scattering} for review of the spinor helicity method for scattering processes of massless particles, the conventions of which, in addition to \cite{Srednicki:2007qs}, we (mostly) adhere to and will not restate. 

Introducing helicity spinors with $SU(2)$ little group structure has consequences for the description of the internal and external structure of scattering amplitudes. Internally, as mentioned above, the starting point is that massive momenta (as representations of the spin group $SL(2,\mathbb{C})$: $p=p^\mu\sigma_\mu$) may be decomposed into two null momenta as
\begin{align}
p^{\dot{\alpha}\beta}=-\sum_I \ra{p_I}^{\dot{\alpha}}\ls{p^I}^{\beta}.\label{spinsum}
\end{align}

The two pairs of left- and right-handed spinors indexed by $I$, $\rs{p_I}$ and $\la{p^I}$, transparently respect an $SU(2)$ symmetry that may be identified with the momentum's little group. These $SU(2)$ indices may be raised and lowered in the usual way to convert between representations and their conjugates:
\begin{align}
\la{p^I}_{\dot{\alpha}}=\varepsilon^{IJ}\la{p_J}_{\dot{\alpha}}\qquad
\rs{p_{I}}_{\alpha}=\varepsilon_{IJ}\rs{p^{J}}_{\alpha}.
\end{align}
Under conjugation, the spinors transform as 
\begin{align}
(\ls{p^I})^\dagger=\ra{p_I}\qquad ({\la{p^I}})^\dagger=-\rs{p_I}.\label{ConjSpinor}
\end{align}
Fundamental tensor representations have lowered indices. We take all scattering states here to be outgoing, so naturally have raised internal indices (including little group) corresponding to the polarisations of the conjugated states.

As usual, $\det(p)=-p^2=m^2$ for mass $m$. As the spinors in (\ref{spinsum}) are conjugates, $\det(p)=\det(\rs{p})\det(\la{p})=|\det(\rs{p})|^2$. The choice of the phase of $\det(\rs{p})$ is free, so $\det(\rs{p})=m$ may be chosen without loss of generality (although see \cite{Plefka:2014fta} for interpretation of the mass and its complex phase as the extra components of a $6$d momentum and its consequences for dual conformal symmetry). The spinors then have bilinear products with themselves
\begin{align}
\da{p^Ip^J}=m\varepsilon^{IJ}\qquad \ds{p^Ip^J}=-m\varepsilon^{IJ},
\end{align}
obey the Weyl equations:
\begin{alignat}{2}
p\rs{p^I}&=-m\ra{p^I}\qquad p\ra{p^I}&&=-m\rs{p^I}\nonumber\\
\ls{p^I}p&=m\la{p^I}\qquad \la{p^I}p&&=m\ls{p^I}\label{Weyl}
\end{alignat}
and the spin sums:
\begin{alignat}{2}
\rs{p_I}_\alpha\ls{p^I}^{\beta}&=m\delta_\alpha^{\beta}\qquad \rs{p_I}_\alpha\la{p^I}_{\dot{\beta}}&&= p_{\alpha\dot{\beta}}\nonumber\\
\ra{p_I}^{\dot{\alpha}}\la{p^I}_{\dot{\beta}}&=-m\delta_{\dot{\beta}}^{\dot{\alpha}}\qquad \ra{p_I}^{\dot{\alpha}}\ls{p^I}^{\beta}&&= -p^{\dot{\alpha}\beta}.\label{SpinSums}
\end{alignat}
The little group index effectively labels the two possible solutions to each of the Weyl equations, which may be rotated into each other by a Wigner rotation.

Externally, the $S$-matrix transforms as a tensor under the little group of each of its external particle legs, being an array of transition matrix entries between states of different spin configurations. An external state of spin $s$ has polarisation wavefunction that can be described by a rank $2s$ symmetric tensor of the little group $SU(2)$. States of a particular polarisation $m_s$ may be extracted from this by choosing the symmetrised component of the tensor with $m_s+s$ indices aligned with the spin direction and $s-m_s$ indices opposite and then normalising. For example, a massive vector particle is described by symmetric polarisation tensor $T^{(I_1I_2)}$, with $m_s=-1,0,1$ states respectively given by $T^{--}$, $\frac{1}{\sqrt{2}}\Big(T^{+-}+T^{-+}\Big)$ and $T^{++}$. See \cite{Georgi:1999wka} for tensor methods to describe spin. We (mostly) restrict to particles of spin $\leq 1$ in this work, although a significant part of the versatility of this formalism is its ability to elegantly describe amplitudes of massive states of any spin.

The possible structures that may appear in the $S$-matrix and are consistent with Lorentz invariance are determined by the number of independent combinations of external state polarisations that can be made. The systematic construction of these was described in \cite{Arkani-Hamed:2017jhn}. Rather than build external polarisations directly from the tensor products of massive spinors (e.g. $T^{(I_1I_2)}=\rs{p^{(I_1}}\rs{p^{I_2)}}$), a direct on-shell construction of elementary amplitudes can be performed instead by using the massive spinors to construct a tensor basis with respect to which the $S$-matrix may be decomposed. Spinors of either chirality (or both) may be used to do this. The coefficients of these basis tensors then represent polarisation-stripped Lorentz tensor amplitudes, in which the possible independent terms may be built out of external momenta and massless spinors. The helicities of massless legs then determine the amplitude's $U(1)$ little group scaling for each massless particle. As a simple example, the $S$-matrix entry for the decay of a massive vector $V_1$ into two massless right-handed fermions $\psi_2$ and $\psi_3$ is determined uniquely by symmetry to be 
\begin{align}
A(V,\psi_2,\psi_3)=g\Big(\ls{1^{(I_1}}^{\alpha_1}\ls{1^{I_2)}}^{\alpha_2}\Big)\times\rs{2}_{\alpha_1}\rs{3}_{\alpha_2}=g\ds{1^{(I_1}2}\ds{1^{I_2)}3},
\end{align}
for some coupling constant $g$. This method of deducing little group structures built out of spinors is used repeatedly throughout this work in constructing superamplitudes. 

Part of the utility of this formalism is that the little group indices are an internal degree of freedom and allow for the polarisation to be projected onto any external spin frame or axis. The procedure for doing this is discussed in \cite{Arkani-Hamed:2017jhn}. In practice, we find that it is clearest to abuse notation and, once such an external frame is specified, simply reinterpret the little group indices as referring to components along this direction. In particular, as it is often most useful, especially in taking massless or high energy limits, to choose spin frames for each particle aligned with their momenta (so that the little group indices simply become helicity indices), we will leave this choice implicit unless stated otherwise. 

In this case, the spinors have massless limits 
\begin{align}
\rs{p^+}\rightarrow \rs{p}\qquad \rs{p^-}\rightarrow 0\nonumber\\
\ra{p^+}\rightarrow 0 \qquad \ra{p^-}\rightarrow -\ra{p}\label{0masslim}
\end{align}
where the spinors without little group indices are the usual spinors for massless momentum $p$. More precisely, the spinors that vanish do so $\mathcal{O}(m)$. The limits may be expressed as 
\begin{align}
\lim_{m\rightarrow 0} \frac{1}{m}\rs{p^-}=\frac{\rs{q}}{\ds{qp}}\qquad\lim_{m\rightarrow 0} \frac{-1}{m}\la{p^+}=\frac{\la{q}}{\da{qp}},\label{1masslim}
\end{align}
where the remaining spinors $\rs{q}$ and $\ra{q}$ become the reference spinors and are ambiguous in the massless amplitude, as their direction arrived in taking the limit is arbitrary (up to requiring $\ds{qp},\da{qp}\neq 0$). In practice, it is often possible to take the limit while avoiding the introduction of the reference by using momentum conservation and other identities. 

For $3$-leg amplitudes involving the factor $x$, the following identities are useful:
\begin{align}
x=\frac{1}{m} \frac{\ls{q}p_2 \ra{3}}{\ds{q3}}=\frac{m\da{q3}}{\la{q}p_2 \rs{3}}
\end{align}
and 
\begin{align}
x\ds{32^I}=\da{32^I}, \quad x\ds{31^I}=-\da{31^I},
\end{align}
\begin{equation}
\frac{\ds{1^I 2^J}}{x} = \frac{\da{1^I 2^J}}{x} + \frac{\ds{1^I 3}\ds{2^J 3}}{m},
\end{equation}
where $p_3$ is the massless leg and $p_1$ and $p_2$ are the massive legs, while $\rs{q}$ and $\ra{q}$ are arbitrary reference spinors, not necessarily related, that satisfy $\ds{q3}\neq 0$ and $\da{q3}\neq 0$.

\subsection{Grassmann Calculus}\label{Grassmann}
The Grassmann variables may be imbued with $SU(2)$ little group indices $\eta_I$. In this case, Grassmann differentiation may be defined in the usual way:
$\frac{\partial}{\partial\eta_{I}}\eta_J=\frac{\partial}{\partial\eta^{J}}\eta^I=\delta_J^I$. However, this requires that the index height on the derivative be raised or lowered with an extra $(-)$ sign: $\frac{\partial}{\partial\eta^{J}}=-\epsilon_{JI}\frac{\partial}{\partial\eta_{I}}$. We note for convenience the identities
\begin{equation}
\half \frac{\partial}{\partial \eta_I} \frac{\partial}{\partial \eta^I} \left( \half \eta_J \eta^J \right) = -1,
\end{equation}
\begin{equation}
\eta_I \eta_J = - \half \epsilon_{IJ} \eta_K \eta^K.
\end{equation}

\noindent The little group invariant Grassmann integration measures are defined here as 
\begin{equation}
\text{d}^2\eta=\frac{1}{2}\epsilon^{IJ}d\eta_I\text{d}\eta_J\qquad\text{d}^2\eta^\dagger=-\frac{1}{2}\epsilon^{IJ}d\eta^{\dagger}_I\text{d}\eta^{\dagger}_J
\end{equation}
where $\int \text{d}\eta_I\eta^{J}=\delta_I^J$ and $\int \text{d}\eta^\dagger_I\eta^{\dagger J}=\delta_I^J$. The index placement on the differential is a property of the differential and not of the variable being integrated - that is, $d\eta_I=d(\eta^I)$. The strange positioning of the index is needed for this operation to be the same as differentiation and is an occurrence of the general topsy-turvyness of Grassmann numbers. Also, as for the derivative, $d\eta^I=-\epsilon^{IJ}d\eta_J$ (and likewise for the conjugate).

The Grassmann Fourier transform of some function $f(\eta)$ of a Grassmann variable $\eta$ is defined as $\tilde{f}$ and these are related by
\begin{align}
\tilde{f}(\eta^\dagger)=\int \text{d}\eta e^{\eta\eta^\dagger} f(\eta)\qquad f(\eta)=\int \text{d}\eta^\dagger e^{-\eta\eta^\dagger} \tilde{f}(\eta^\dagger).
\end{align}
The Fourier transform from the $\eta^\dagger$ basis to the $\eta$ basis in $\mathcal{N}=1$ is effected by the replacements
\begin{equation}
1 \rightarrow - \half \eta_{I}\eta^I, \qquad \eta^{\dagger I} \rightarrow \eta^I, \qquad - \half \eta^{\dagger I} \eta^\dagger_I \rightarrow 1.
\end{equation}

\noindent For multiplets without a central charge, the Grassmann variables have massless limits in the helicity basis
\begin{equation}
\eta_- \rightarrow \eta, \qquad \eta_+ \rightarrow \hat{\eta}.
\end{equation}
Here, $\hat{\eta}$ represents the redundant variable left-over from the division of the massive multiplets into smaller massless multiplets that each represent the smaller massless SUSY algebra. For the exceptional case of BPS multiplets, $\hat{\eta}=\eta^\dagger$, while for anti-BPS multiplets the limit picks up an extra negative sign.

\section{Comments on Higher-leg Amplitudes in SQCD}\label{higherLeg}
We here make some comments on various massive quark and multigluon amplitudes and rederive them in the little group covariant notation. Again, the arguments presented here are parallel to \cite{Boels:2011zz} and \cite{ArkaniHamed:2008gz}. Amplitudes stated here will be colour-stripped partial amplitudes, following the usual rules for Yang-Mills theories, as prescribed in e.g. \cite{Mangano:1990by}. This discussion is supplementary to further comments made in \cite{Herderschee:2019dmc} concerning the relation between (S)QCD amplitudes and Coulomb branch amplitudes of $\mathcal{N}=4$ SYM.

Firstly, the supersymmetric Ward identities provide relations and constraints between component amplitudes that can be exploited. Supersymmetry transformations can be found that set a Grassmann generator for a particular leg to $0$. In particular, under the action of $-\ds{\theta Q}$, $\eta_{j,I}$ is translated to $\eta_{j,I}-i\ds{\theta j_I}$ for each leg $j$ (if the leg is massless, just omit the little group index). This can be used to set $\eta_{i,I}=0$, for some single leg $i$ with polarisation in some direction given by $I$ in some little group frame, by choosing $\ls{\theta}=\frac{-i}{m}\eta_{i,I}\ls{i^I}+C\ls{i_I}$ (no sum over $I$ is implied). Here, $C$ represents the remaining unused degree of freedom in the supersymmetry parameter. Component amplitudes that are obtained by integrating the superamplitude in the Grassmann parameters that are translated are unaffected by this transformation, because the integration variable can be likewise translated. After changing variables to absorb the supertranslation, the resulting integrand is completely independent of $\eta_{i,I}$, so integrating over it will give $0$. The component amplitudes obtained by such projections must therefore be $0$ by supersymmetry. 

Simple illustrative examples of this are the squark-antisquark and $n$-gaugino amplitude and the squark-antisquark $n$-gluon amplitude. 
\begin{align}
A[\overline{\widetilde{Q}}_R\lambda^+\cdots\lambda^+\widetilde{Q}_L]=\int \text{d}^2\eta_1 \prod_i\int \text{d}\eta_i \int \text{d}^2\eta_{n+2} \mathcal{A}[\overline{\mathcal{Q}},G^{+}\ldots G^{+},\mathcal{Q}]=0\\
A[\overline{\widetilde{Q}}_Rg^-\cdots g^-\widetilde{Q}_L]=\int \text{d}^2\eta_1 \prod_i\int \text{d}\eta_i \int \text{d}^2\eta_{n+2}\mathcal{A}[\overline{\mathcal{Q}},G^{-}\ldots G^{-},\mathcal{Q}]=0
\end{align}
Identical arguments in the $\eta^\dagger$ basis may be used to show that the $CP$-conjugate amplitudes are also $0$. Identical arguments also demonstrate that amplitudes with additional squark and antisquark pairs $(\overline{\widetilde{Q}}_R,\widetilde{Q}_L)$ are $0$.

Vanishing amplitudes of massive quarks, states of non-trivial polarisation, may also be obtained similarly. If all of the little group axes of the quarks are aligned, then the transformation $\eta_{j,J}-i\ds{\theta j_J}$ does not affect the Grassmann numbers with opposite spin components to $J$ - which is now the same direction for each massive field. Thus the superamplitude integrated over only these components will be independent of the Grassmann variable that is being eliminated, so must vanish. This derives the fact that amplitudes with quarks and antiquarks all of identical polarisation are $0$, as well as those that include some number of gluons or gauginos of identical helicity. These include amplitudes that are inherited by pure QCD at tree-level. The argument can be easily combined with that used in the previous paragraph to extend these vanishing amplitudes to those involving squarks.

The extra degree of freedom in the supersymmetry parameter can be further utilised to derive the vanishing of a further class of amplitudes. Choosing $C=\frac{-i}{m}\frac{\ds{i^Ij_K}}{\ds{i_Ij_K}}\eta_{i,I}$ (if $j$ is massless, omit its little group index in this expression), the Grassmann variable for leg $j$ does not shift under the supersymmetry transformation performed in the examples above. This means that the variables $\eta_{j,K}$ need not be integrated in order to obtain a vanishing amplitude. Thus one extra particle in any spin state may be added to any of the amplitudes above and the result will still be $0$.

Tree-level amplitudes involving a quark-antiquark pair and any number of gluons of the same helicity have been previously determined in \cite{Schwinn:2006ca, Schwinn:2007ee} and little group covariantised in \cite{Ochirov:2018uyq}. A compact expression exists that may be derived inductively using BCFW recursion by shifting the massless legs in the usual way \cite{Badger:2005zh}. The superamplitudes to which these amplitudes belong have the interesting property that they are fully determined by a single component amplitude, which we show in Appendix B of \cite{Herderschee:2019dmc} by projecting these superamplitudes out from the massive $\mathcal{N}=4$ theory.

\bibliography{superspace}{}
\bibliographystyle{JHEP}
\end{document}